\documentclass[11pt]{amsart}

\usepackage{xcolor}
\newcommand{\change}[1]{{#1}}

\usepackage{xspace}

\usepackage[utf8]{inputenc}

\usepackage{amssymb,amsmath,amsxtra,stmaryrd,verbatim,framed,wasysym}

\usepackage[psamsfonts]{eucal}

\usepackage{graphics}

 \newtheorem{Theorem}{Theorem}
 
 \newtheorem{Proposition}[Theorem]{Proposition}
 \newtheorem{Lemma}[Theorem]{Lemma}

 \newtheorem{Definition}[Theorem]{Definition}

 \theoremstyle{Remark}
 \newtheorem{Remark}[Theorem]{Remark}

 \newtheorem{Property}[Theorem]{Property}

\newcommand{\varletter}[1]{\mathcal{#1}}
\newcommand{\Q}{\varletter{Q}}

\newcommand{\K}{\varletter{K}}
\newcommand{\B}{\varletter{B}}
\newcommand{\V}{\varletter{V}}
\newcommand{\varP}{\varletter{P}} 

\newcommand{\dletter}[1]{\mathbb{#1}}
\newcommand{\DD}{\dletter{D}} 
\newcommand{\ZZ}{\dletter{Z}}
\newcommand{\CC}{\dletter{C}}  
\newcommand{\NN}{\dletter{N}} 
\newcommand{\PP}{\dletter{P}} 

\newcommand{\norm}[1]{\Vert#1\Vert}
\newcommand{\inprod}[2]{\langle #1, #2\rangle}
\newcommand{\kinprod}[2]{\langle #1\mid #2 \rangle}

\newcommand{\ket}[1]{\vert #1\rangle} 
\newcommand{\tek}[1]{\langle #1\vert} 
\newcommand{\cb}[1]{\mathsf{CB}(#1)}
\newcommand{\SPAN}[1]{\mathsf{span}(#1)}
\newcommand{\CoC}{\widetilde{\CC}}

\newcommand{\CConf}{\mathfrak{C}}
\newcommand{\GConf}{\mathfrak{C}}
\newcommand{\IConf}{\GConf^{init}}
\newcommand{\FConf}{\GConf^{fin}}
\newcommand{\SConf}[1]{\GConf^{#1}}

\newcommand{\nstring}[1]{\underline{#1}}

\newcommand{\lub}{\bigsqcup}

\newcommand{\BeV}{B\&\!V\xspace}

\newcommand{\val}{\mathbf{val}}

\newcommand{\pd}[1]{\mathbf{P}_{#1}}
\newcommand{\prob}[2]{\pd{#1}(#2)}

\newcommand{\outobs}[3]{\ket{#1}\downarrow_{#2}\ket{#3}}
\newcommand{\obsout}[2]{#1\downarrow_{#2}}
\newcommand{\PR}{\mathsf{Pr}}
\newcommand{\pr}[1]{\PR\{#1\}}

\newcommand{\abextra}[1]{\overline{#1}}

\newcommand{\symextra}[1]{\overline{#1}}

\newcommand{\dind}[2]{\genfrac{}{}{0pt}{1}{#1}{#2}}

\newcommand{\Bev}{Bernstein \& Vazirani}

\title{Quantum Turing Machines:\\ computations and  measurements}

\date{\today}

\author[S. Guerrini]{Stefano Guerrini} \address{Stefano Guerrini\\
  LIPN, UMR 7030 CNRS, Institut Galilée, Université Paris13, Sorbonne
  Paris Cité} 

\email{stefano.guerrini@univ-paris13.fr}

\author[S. Martini]{Simone Martini$^{\S}$} \address{Simone Martini\\
  Dipartimento di Informatica -- Scienza e Ingegneria, Università di
  Bologna, and Inria Sophia-Antipolis}

\email{simone.martini@unibo.it} 
\thanks{$^{\S}$ Research partially conducted while on sabbatical leave at the Collegium -- Lyon Institute for Advanced Studies, 2018-2019; partial support from INdAM-GNSAGA}

\author[A. Masini]{Andrea Masini}

\address{Andrea Masini\\ Dipartimento di Informatica, Università di Verona}

\email{andrea.masini@univr.it}

\begin{document}

\begin{abstract}%
  Contrary to the classical case, the relation between quantum
  programming languages and quantum Turing Machines (QTM) has not
  being fully investigated. In particular, there are features of QTMs
  that have not been exploited, a notable example being the intrinsic
  infinite nature of any quantum computation. In this paper we propose
  a definition of QTM, which extends and unifies the notions of
  Deutsch and Bernstein \& Vazirani. In particular, we allow both
  arbitrary quantum input, and meaningful superpositions of
  computations, where some of them are ``terminated'' with an
  ``output'', while others are not. For some infinite computations an
  ``output'' is obtained as a limit of finite portions of the
  computation. We propose a natural and robust observation protocol
  for our QTMs, that does not modify the probability of the possible
  outcomes of the machines. Finally, we use QTMs to define a class of
  quantum computable functions---any such function is a mapping from a
  general quantum state to a probability distribution of natural
  numbers. We expect that our class of functions, when restricted to
  classical input-output, will be not different from the set of the
  recursive functions.\\ \ \\
  {\sc Keywords}: Quantum Turing Machines; Computability theory; Computer
  science; Theory of programming languages.\\ \ \\
  \emph{To appear on Applied Sciences, MDPI}
\end{abstract}

\maketitle


\section{Introduction}
\label{sec:introduction}

It is an early recognition of computer science that programming (and programming languages) could benefit from solid semantical foundations. At the end of the 50s of last century, both for clarity purposes and for the need of an entrance ticket into the ``science club'', some prominent scientists started using mathematical techniques for the \emph{description} of programming languages (e.g., Chomsky's generative grammars, in the form of Backus-Naur Form~\cite{Backus1959-BNF}). Even more important is the proposal of general semantical models for the \emph{meaning} of programs, represented at its best by McCarthy's ``mathematical theory of computation''~\cite{McCarthy1963}, where a definition of the class of computable functions is presented as a general model for the  semantics of (Algol) programs. The model is clearly idealised, in which it assumes that execution happens without physical  limitations of space and time, in what will be, from that moment onwards, the \emph{standard model} for programming languages, where one may assume Turing-completeness, despite each implementation being a finite state machine. The standard model builds on a long process during which Turing machines are viewed as idealised models of actual computers, and thus provide a basis for some theoretical investigation, among which the first studies on computational complexity. 

Quantum computations and quantum programming languages have followed a different path. A solid and ever growing body of research is present on semantics and mathematical methods \change{%
(e.g., see \cite{DLMZentcs11,DLZorzi15,Zorzi16,iandcVZ17, PRZLlinearity18,axiomsMZ19,jarPPZ19})
}, but the relations of these languages with the basic notion of (quantum) Turing machines (as introduced by Deutsch \change{in his seminal paper on quantum computation~\cite{Deu85}; see also~\cite{Be80} for a quantum-mechanical model of Turing machines}) has received much less study than in the classical case.  The exception is, of course, the  definition of quantum Turing machines (QTMs) by Bernstein \& Vazirani~\cite{BerVa97} (\BeV, from now on) and its use in the establishment of a sound quantum computational complexity. Their machines, however, have several strong constraints,  reasonable for the purpose of a theory of computational complexity, but which do not permit to exploit all the features of the idealised model of computation that QTMs provide. In particular, meaningful computations are defined only for classical inputs (a
single natural number with amplitude 1), and \BeV's ``well-behaved'' QTMs
``terminate'' synchronously---either all paths in superpositions
enter a final state at the same time, or all of them diverge. As a
consequence, there is no chance to study---and give meaning---to
infinite computations, since for \BeV's ``well-behaved'' QTMs ``non termination'' corresponds to classical divergence.

Our aim is to provide a definition of QTMs with both quantum data and quantum control, hence broader than \BeV's one, and to base on this notion an explicit definition of a class of functions ``computed by QTMs''. First of all, we allow a QTM to ``start'' its computation on an arbitrary quantum input---that is, a
denumerable superposition of  configurations, each representing a natural number. Second, since any computation of a QTM is always an infinite one, we want to give meaning also to some of these infinite computations. For such infinite
behaviours, an ``\emph{output}'' is obtained as a limit of finite portions
of the computation. In order to accommodate these two aims with the unitary condition of the evolution, we have first to carefully define the very notion of QTM, in such a way that, when a  computation of the machine enters into a ``final
state'', its evolution continues to remain in that final state,
without changing the output written on the tape. This kind of
evolution is obtained by enriching the machines by means of a suitable
\emph{counter} which plays no role during the standard evolution of
the machine, but starts to be increased when a path of the computation enters
into a final state. (A dual treatment, by reversibility, is done for the initial state.)
Differently from \BeV, any computation path (that is, any element of the superposition)
evolves independently from the others: any path may terminate at its
own time, or may diverge.
Like \BeV, we give local conditions ensuring that the resulting evolution respects the unitary condition.

We propose a suitable protocol for termination\footnote{\change{In all the paper, with the exception of Section~\ref{ssec:readout-problem} when discussing papers of other authors, we use ``termination'' and not ``halting.'' Since there is no real halting for QTMs, we use a term different from the standard one used for classical TMs, to remark the difference.}} and observation of the result, which is robust with respect to the choice of the time instants in which the measurements for observing the results are made. This is in contrast to both \BeV, where one has to know the exact termination time, and to 
Deutsch, where a partial measurement is made at every single step of the computation. The termination criteria for \BeV{} and Deutsch is also one of the main obstacles for their use to define meaningful infinite computations. 

Once we have a robust notion of QTMs and their observation protocol, we define the notion of ``function computable by a
QTM,'' as a mapping between superpositions of
initial classical configurations to
probability distributions of natural numbers, which is obtained (in
general) as a limit of an infinite QTM computation. Since not all infinite computations are meaningful to the limit, the resulting class of quantum computable functions contains also non-total functions, as it is usual in computability theory. 

We stress that we are not proposing a kind of ``super Turing computability''. We expect that our class of functions, when restricted to classical input-output \change{and computable amplitudes}, will be not different from the set of the recursive functions \change{(see the end of Section~\ref{ssec:quant-part-comp-fun})}. 

We believe interesting to have a class of machines and functions which could be seen as a simple, and standard, model of quantum programs, thus opening the
door to a quantum computability theory which could be of use for the semantics of quantum programming languages.

\subsection*{Organization of the paper} 

Section~\ref{sec:qtm} introduces the central notion of Quantum Turing Machines; the technical definitions are preceded by a lengthy discussion on the motivations of the most delicate issues. We conclude the section with a theorem giving local conditions for unitarity, as in \cite{BerVa97}. 
Section~\ref{sec:quant-comp-fun} uses QTMs to define a class of partial quantum computable functions.
Section~\ref{sec:observables}  deals with the problem of observables with respect to QTM computations. The protocols of Deutsch~\cite{Deu85} and ~\cite{BerVa97} are discussed, and a new observational  protocol is proposed, proving that it agrees with the notion of computable functions of the previous section. In particular, we discuss in details Deutsch's protocol, in which sense our proposal is different, and how it is related to Ozawa's termination protocol~\cite{Ozawa-PRL-98}.
In Section~\ref{sec:comp-BandV-2} we propose a detailed comparison with the approach of \BeV{}. 
Section~\ref{sect:related} is a detailed  discussion of related work and other approaches. Finally, 
we discuss in the Appendix~\ref{sec:HS} some additional technical issues.

%

\section{Quantum Turing Machines}
\label{sec:qtm}

In this section we define quantum Turing machines.  
Before the formal statements (Definitions~\ref{def:pQTM},
\ref{Def:config}, \ref{Def:quantum-config}, etc.) we will discuss in
detail some of the notions that will be introduced, to motivate and
explain our technical choices. We assume the reader be familiar with
classical Turing machines (otherwise, see~\cite{Davis58}), 
\change{which we assume defined with a tape infinite in both directions.}

Like a classical Turing Machine (TM), a Quantum Turing Machine (QTM)
has a tape (a sequence of cells) containing symbols (one in each cell)
from a finite \emph{tape alphabet} $\Sigma$, which includes at least
the symbols $1$ and $\Box$: $1$ is used to code natural numbers in
unary notation, while $\Box$ is the blank symbol. We shall consider
computations starting from tapes containing a sequence of $n+1$
symbols $1$ (the encoding of the natural number $n$); thus, in the
following, for any $n\in\NN$, we shall use $\nstring{n}$ to denote the
string $1^{n+1}$. By the Greek letters $\alpha, \beta$, possibly
indexed, we shall instead denote strings in $\Sigma^*$, and by
$\alpha\beta$ we shall denote the concatenation of $\alpha$ and
$\beta$. Finally, we shall use $\lambda$ to denote the empty string.

\subsection{Plain configurations} 
\label{ssec:plain-config}

The basic elements to describe the configuration of a QTM are the
finite sequences of symbols on the tape (as usual, we assume that only
a finite portion of the tape contains non-blank symbols), the current
internal state of the machine, and the current position of the head
reading a symbol in a cell of the tape. We assume that each cell of
the tape has a fixed address (an integer number), corresponding to its
position on the tape. That is, we assume the tape to be a function
$T:\ZZ\to\Sigma$, such that, at any moment of the execution,
  only a finite, contiguous part of the tape is not empty, i.e., at
  any step of the execution, there are two constants $h \leq k$ s.t.,
  $T(i) = \Box$ for $i < h$ and $i \geq k$ (thus, when $h=k$ the tape
  is empty).

A \emph{canonical plain configuration} of a given QTM $M$
is a quadruple
$\langle\alpha,q,\beta, i\rangle\in\Sigma^*\times
Q\times\Sigma^*\times\ZZ$, s.t.:
\begin{enumerate} 
\item $q\in Q$ is the \emph{current state}, where $Q$ is the finite
  set of the internal states of $M$.
  
\item
    $\beta\in\Sigma^*$ is the right content of the
    tape (w.r.t.\ the head position), where $\Sigma$ is the tape
    alphabet of the machine $M$. If $\beta=u\beta'\neq\lambda$, the
    symbol $u\in\Sigma$  is
    the \emph{current symbol}, that is
    the content of the \emph{current cell}, while $\beta'\in\Sigma^*$ is the
    longest string on the tape ending with a symbol different from
    $\Box$ and whose first symbol (if any) is written in the cell
    immediately to the right of the current cell. When $\beta=\lambda$ instead,
    the right content of the tape, including the current cell, is empty, and 
    the current symbol is $\Box$.

\item $\alpha$ is the left content of the tape (w.r.t.\ 
  the head position). That is, it is either the empty string
  $\lambda$, or it is the longest
  string on the tape starting with a symbol different from $\Box$, and
  whose last symbol is written in the cell immediately to the left of
  the current cell.
\item $i$ is the address of the current cell.
\end{enumerate}
%

Given $\alpha, \beta\in\Sigma^*$, let
$\alpha=a_0a_1\ldots a_{u-1}$ and $\beta=b_0b_1\ldots b_{v-1}$, with
$u,v\geq 0$. The plain configuration $\langle\alpha,q,\beta, i\rangle$
corresponds to the tape
\begin{equation*}
  T(j) =
  \begin{cases}
    a_{j-i+u} & \mbox{for $i-u \leq j < i$} \\
    b_{j-i} & \mbox{for $i \leq j < i+v$} \\
    \Box & \mbox{otherwise}
  \end{cases}
\end{equation*}
where $T(j)\neq \Box$, \change{implies that} $i-u \leq j < i+v$.

In a canonical plain configuration
$\langle \alpha, q, \beta, i\rangle$ the string $\alpha$ does not
start with a $\Box$, and $\beta$ does not end with a $\Box$. This
ensures that each tape corresponds to a unique canonical plain
configuration, and vice versa. However, since it will be useful to
manipulate configurations which are extended with blank cells to the
right (of the right content) or to the left (of the left content),
  we shall also consider plain configurations
  $\langle \alpha, q, \beta, i\rangle$ in which there is no
  restriction on $\alpha$ and $\beta$ and we 
    equate them up to the three equivalence
relations induced by the following equations
\begin{gather*}
  \alpha \simeq_l \Box\alpha
  \qquad\qquad\beta \simeq_r \beta\Box\\
  \langle \alpha, q, \beta, i\rangle \simeq 
  \langle \alpha', q, \beta', i\rangle \qquad\qquad 
  \mbox{when $\alpha\simeq_l\alpha'$ 
    and $\beta\simeq_r\beta'$}
\end{gather*}

  It is readily seen that any plain configuration is
  equivalent to a unique canonical plain configuration, and that
  two configurations are equivalent iff they are in the same state,
  the address of the head is the same,
  and they correspond to the same tape.

\subsection{Hilbert space of configurations}
\label{ssec:hilb-space-conf}
We will see in the following that a quantum configuration of a QTM cannot be simply a plain configuration, but it must be a weighted superposition of configurations: a vector
of the Hilbert space $\ell^2({\mathcal C})$, where ${\mathcal C}$ is a 
set of  configurations, like the plain ones defined above.

We recall that, for any denumerable set $\B$, $\ell^2(\B)$ is the
Hilbert space of square summable $\B$-indexed sequences of complex
numbers
\[
  \left\{ %
    \phi:\B \rightarrow\CC \mid \sum_{C\in \B}|\phi(C)|^2  < \infty 
  \right\} %
\]
equipped with an \emph{inner product} $\kinprod{.}{.}$ and the
\emph{euclidean norm} $\norm{\phi}=\sqrt{\kinprod{\phi}{\phi}}$, and
that $\ell_1^2(\B)$ denotes the set of vectors
$\{\phi \mid \phi\in \ell^2(\B) \ \&\ \norm{\phi}=1\}$.

By using Dirac notation, we shall write $\ket{\phi}$ to denote the
vector of $\ell^2(B)$ corresponding to the function
$\phi:\B \to \CC$.  Moreover, for every $C\in \B$, we shall write
$\ket{C}$ to denote the vector corresponding to $C$, that is, the
function equal to $1$ on $C$, and equal to $0$ on the other elements
of $\B$. Finally, we remark that any vector $\ket{\phi}$ of $\ell^2(\B)$
can be written as $\sum_{i\in I} d_i \ket{C_i}$, for some denumerable
set of indexes $I$ s.t.\ $\{C_i\mid i\in I\}\subseteq \B$ and
$d_i\in\CC$, for every $i\in I$. For more details on the basic notions of
Hilbert spaces, see Appendix~\ref{sec:HS}.

\subsection{Transitions of a QTM}
\label{ssec:transitions-qtm}

Given a quantum configuration $\phi$ of a QTM $M$ (see Section~\ref{ssec:q-configurations} 
for the formal definition), the main idea
introduced by Deutsch~\cite{Deu85} is that the machine evolves into
another quantum configuration $\ket{\psi}=U \ket{\phi}$, where $U$ is
a unitary operator on the Hilbert space of the configurations of $M$.
By linearity and continuity, this also means that, if ${\mathcal C}$ is the
set of  configurations on which we define the Hilbert space of
quantum configuration $\ell^2({\mathcal C})$ of $M$, the operator $U$, and
then the behaviour of $M$, is completely determined by the value of
$U$ on the elements of ${\mathcal C}$. \BeV~\cite{BerVa97} refines this
point by assuming that, as in a classical TM, for every $C\in {\mathcal C}$,
the transition from $\ket{C}$ to a new configuration $U\ket{C}$
depends only on the current state of $M$ and on the current symbol of
$C$, and that $U\ket{C}$ is formed of configurations obtained by
replacing the current symbol $u$ of $C$ with a new one, and by moving
the tape head to the left or the right. Therefore, if we denote by
$\DD=\{L,R\}$ the set of the possible movements on the tape (where $L$
stands for left and $R$ for right), and $q$ is the current state of
$M$, we have
$$U(\ket{C}) = \sum_{p,v,d}
\delta(q,u)(p,v,d)\,\ket{C_{p,v,d}}
$$
where $p$ ranges over the states of $M$, $v$ ranges over its tape
alphabet, $\delta(q,u)(p,v,d)\in\CC$, and $C_{p,v,d}$ is the new
configuration obtained from $C$ by changing the current state from $q$
to $p$, by replacing the current symbol $u$ with $v$, and by moving
the tape head in the direction $d$.

\subsection{Initial and final configurations}
\label{ssec:init-final-conf}

In \BeV, the transition rule described above applies to every state of
the machine. However, as already remarked in the introduction, this
implies a severe restriction on the machines that one can actually
consider as valid ones---the reversibility of unitary operators, and
the problem of how to read the output, force to ask that, in a
``well-behaved'' QTM, if a configuration in superposition enters the
final state, then all other configurations of the superposition must
also enter into the final state.

The key point is that a QTM cannot stop into some final configuration
$C_f$, since the unitarity of $U$ requires that $U(\ket{C_f})$ be
defined. On the other hand, we cannot assume that, after reaching some
final configuration $C_f$, the computation loops on it neither. As this
would imply that $C_f$ could not be reached from any other
configuration $D\in{\mathcal C}$: if $U(\ket{C_f})=\ket{C_f}$, then
$\ket{C_f}=U^{-1}(\ket{C_f})$, and
$U(\ket{D})(C_f)= \kinprod{U(\ket{D})}{C_f}=
\kinprod{D}{U^{-1}(\ket{C_f})}= \kinprod{D}{C_f}=0$, for $D\neq C_f$.
When a final configuration $C_f$ is reached, the machine must evolve
into a different configuration which preserves the output written on
the tape, without breaking the unitarity of the evolution operator
$U$. Let assume to have a denumerable set of indexed configurations
$\langle C_f, n\rangle$ obtained by adding a \emph{counter} $n\in\NN$
to $C_f$, for every final configuration $C_f$. The configuration
$\langle C_f, 0 \rangle$ plays the usual role of the plain
configuration $C_f$ and it is the only one that can be reached from a
non final configuration; for every $n\in\NN$, the configuration
$\langle C_f, n+1\rangle$ is instead the only successor of
$\langle C_f, n\rangle$, that is, $U(\ket{C_f, n}) = \ket{C_f,n+1}$,
where $\ket{C_f,n}$ is the base vector corresponding to
$\langle C_f, n\rangle$. In other words, the counter $n$ is
initialised to $0$ and plays no role while the state is not final;
when a branch of the computation enters a final state, the counter is
still at $0$, but it is increased by $1$ at each following step.

Final configurations are not the only configurations on which it would
be useful to loop. In some cases, in order to assure that the
operator $U$ is unitary, we need to introduce additional target
configurations that behave as sinks, from which the machine cannot
get out \change{(see Remark~\ref{rem:source-target}, and} the example in Subsection~\ref{ssec:ex:identity}, whose
graph representation is given in Figure~\ref{fig:ide}). We have then
to consider a whole set $\Q_t$ of target states, containing the final
state $q_f$, s.t.\ for every plain configuration
$C_t = \langle\alpha,q_t,\beta, i\rangle$, with $q_t\in\Q_t$, we have a
set of indexed configurations $\langle C_t, n \rangle$, for $n\in\NN$,
s.t.\ $U(\ket{C_t, n}) = \ket{C_t,n+1}$.

Moreover, when $U$ is unitary, its inverse $U^{-1}$
is unitary too, and can be applied to any initial
configuration. Therefore, even if we assume that a computation always
starts from some initial plain configuration
$C_j=\langle\alpha,q_j,\beta,i\rangle$, such a $C_j$ must have a
predecessor $\ket{C_{j,1}}=U^{-1}(\ket{C_j})$, and more generally a
$n$-predecessor $\ket{C_{j,n}}=U^{-n}(\ket{C_j})$. As already done for final
configurations, we associate to every $C_j$ a set of indexed
configurations $\langle C_j, n\rangle$, with $n\in\NN$, s.t.\
$U^{-1}(\ket{C_j, n}) = \ket{C_j,n+1}$, for $n \in\NN$.  For $n > 0$,
we get then $U(\ket{C_j, n}) = \ket{C_j,n-1}$; while $U(\ket{C_j, 0})$
has the usual behaviour expected from the machine on the plain initial
configuration $C_j$. As for final state and final configurations, it
is useful to generalise the above behaviour to a set of source
configurations corresponding to a set $\Q_s$ of source states which
contains the initial state $q_j$.

\subsection{Pre Quantum Turing Machines}
\label{ssec:pre-quantum-turing}

In order to formally define QTMs, we first define  a notion of pre
Quantum Turing Machine, or pQTM. As we shall see later
(Definition~\ref{def:qtm}), a QTM is a pQTM whose evolution operator
is unitary. Since the behaviour on target states and on source states
whose counter is greater than $0$ is fixed, in order to completely
describe a pQTM, it suffices to give a transition function
$\delta_0 : ((\Q\setminus \Q_t)\times\Sigma) \to \ell_1^2((\Q\setminus
\Q_s) \times \Sigma \times \DD)$ which, for every configuration $C$
with current state $q$ and current symbol $u$, gives the weight
$\delta_0(q,u)(p,v,d)$ of $\ket{C_{p,v,d}}$ in the superposed
configuration reached from $\ket{C}$, where $C_{p,v,d}$ is obtained by
replacing $v$ for $u$, by moving the tape in the $d$ direction, and by changing the current state to $p$.

\begin{Definition}[Pre Quantum Turing Machine]\label{def:pQTM}
  Given a finite set of states $\Q$ and an alphabet $\Sigma$, a  \emph{pre Quantum
  Turing Machine} (pQTM) is a tuple
  $$M=\langle \Sigma, \Q, \Q_s, \Q_t, \delta_0, q_i, q_f\rangle$$
  where
  \begin{itemize}
  \item $\Q_s\subseteq \Q$ is the set of \emph{source states} of $M$, and
    ${q_i}\in \Q_s$ is a distinguished source state named the \emph{initial
      state} of $M$;
  \item $\Q_t\subseteq \Q$ is the set of \emph{target states} of $M$, and
     $q_f \in \Q_t$ is a distinguished target state named the
    \emph{final state} of $M$;
  \item $\Q_s$ and $\Q_t$ have the same cardinality and
    $\Q_s \cap \Q_t =\emptyset$;
  \item
    $\delta_0 : ((\Q\setminus \Q_t)\times\Sigma) \to
    \ell_1^2((\Q\setminus \Q_s) \times \Sigma \times \DD)$ is the
    \emph{quantum transition function} of $M$, where $\DD=\{L,R\}$.
  \end{itemize}
\end{Definition}

\change{%
\begin{Remark}[Source and target states]\label{rem:source-target}
  Source and target states must be introduced to complete a QTM when
  the transitions implementing its expected behaviour fail to assure
  the local conditions of Theorem~\ref{theor:local}, required to get a
  unitary time evolution operator. In particular, the first condition
  states that the norm of the vector $\delta(q,u)$ corresponding to
  the transition from a configuration $C$ with current state $q$ and
  current symbol $u$, to a superposed configuration
  $\sum\delta(q,u)(p,v,d) \ket{C_{p,v,d}}$ must be equal to 1. Anyhow,
  in some case, the norm of the vector of the weights implementing the
  expected behavior of the machine might give a value smaller than 1,
  while in no case it cannot exceed the value of 1. To achieve the
  required local condition, we can add some new transitions to some
  special target state acting as a sink (i.e., from which the machine
  cannot get out), with a procedure reminding the one used to complete
  a non-deterministic automata by connecting all the omitted
  transitions to a sink state. The main difference is that in the QTM
  case more than one target state may be necessary to achieve the
  goal. Moreover, because of the reversibility of QTMs, a similar
  condition must hold for the transitions entering into a given state. So,
  we must also take into account the case in which we need to
  complete the implementation of the expected behaviour by adding new
  transitions from some source states. Finally, we remark that the
  requirement that the sets of source and target states have the same
  cardinality is used to prove the unitarity of the evolution
  operator associated to our QTMs via a reduction to the \BeV QTMs
  (see Theorem~\ref{theor:local}). We do have another,
  independent, and direct proof of this result (that we omit in
  this paper), and which also exploits the equinumerosity of source 
  and target states. We conjecture then that this is
  indeed a necessary condition.
\end{Remark}
}

\begin{Remark}[On  g\"odelization of QTMs]
  For a definition of gödelizable QTMs,
  we must restrict the definition of the transition function
  to computable complex numbers \change{(see also~\cite{BerVa97}.)}
  
  \begin{Definition}[computable numbers]
    A real number $x$ is computable if there exists a deterministic
    Turing machine that on input $1^n$ computes a binary
    representation of an integer
    $m\in\ZZ$ such that $|\frac{m}{2^n} - x|\leq \frac{1}{2^n}$.
    \\
    The set $\CoC$ of the computable complex numbers is the set of the
    complex numbers whose real and imaginary parts are both
    computable.
  \end{Definition}

  Now, let $\tilde{\ell} _1^2{(\B)}$ be the subset of
  ${\ell}_1^2{(\B)}$ s.t.\ the coefficients of the vectors are bound
  to be in $\CoC$. We can say that a transition function
  $\tilde{\delta}_0$ is computable when
  \[
    \tilde{\delta}_0 : ((\Q\setminus \Q_t)\times\Sigma) \to
    \tilde{\ell_1^2}((\Q\setminus \Q_s) \times \Sigma \times \DD).
  \]
  The gödelization of a QTM whose transition function is computable
  proceeds in the usual way. First of all, just observe that
  ${\ell_1^2}((\Q\setminus \Q_s) \times \Sigma \times \DD)$ is a
  finite dimensional space, then that
  each complex number $x+i y\in \CoC$ involved in the gödelization of
  $\tilde\delta_0$ can be replaced by a pair $(n_x, n_y)$, where $n_x$
  and $n_y$ are the Gödel numbers of the Turing machines computing
  $x$ and $y$, respectively.
\end{Remark}

\subsection{Configurations} \label{ssec:configurations}

We have already seen that, in order to properly deal with source and
target states, we have to associate a counter to those states. For the sake of
uniformity, we shall add a counter to every plain configuration. 
The counter will always be $0$ for every non-source or non-target configuration.

\begin{Remark}[The counter]
  The counter can be seen as an additional device (for instance, as an
  additional tape or as a counting register) or directly implemented
  by a suitable extension of the basic Turing machine (for instance,
  by extending the tape alphabet). None of these implementations is
  canonical or has a direct influence in what will be presented in the
  following. 
  A more detailed discussion of the
  implementation of the counter is given in
  Appendix~\ref{sec:impl-counter}.
\end{Remark}

\begin{Definition}[configurations]\label{Def:config}
  Let $M=(\Sigma, \Q, \Q_s, \Q_t,\delta_0, q_0, q_f)$ be a pQTM. A
  \emph{configuration} of $M$ is a quintuple
  $\langle\alpha,q,\beta,i, n\rangle \in {\Sigma}^*\times \Q\times
  {\Sigma}^*\times \ZZ \times \NN$,
  where $\langle \alpha,q,\beta,i\rangle$ is a
  plain configuration, and $n$ is a \emph{counter} associated to the
  configuration s.t.\ $n=0$, when $q \not\in \Q_s \cup \Q_t$.  A
  configuration of $M$ is a \emph{source}/\emph{target configuration}
  when the corresponding state is a source/target state, and it is a
  \emph{final}/\emph{initial configuration} when the current state is
  final/initial. We have the following notations:
  \begin{itemize}
  \item $\GConf_M$ is the set of the configurations of $M$.
  \item $\SConf{s}_M$ and $\SConf{t}_M$ are the sets of the source and
    of the target configurations of $M$, respectively.
  \item $\IConf_M$ and $\FConf_M$ are the sets of the initial and
    final configurations of $M$, respectively.
  \item $\SConf{0}_M$ is the set of the configurations
    $\langle \alpha, q, \beta, i, 0\rangle$ of $M$.
  \end{itemize}
\end{Definition}

In the following, the subscript $M$ in $\GConf_M$ and in the other names
indexed by the machine may be dropped when clear from the context.


\subsection{Quantum configurations}
\label{ssec:q-configurations}

The evolution of a pQTM is described by superpositions of
configurations in $\GConf_M$.  If $\B \subseteq \CConf_{M}$ is a set
of configurations, a superposition of configurations in $\B$ is a
vector of the Hilbert space $\ell^2(\B)$ (see,
e.g.,~\cite{Con90,RomanBook}).  Quantum configurations of a pQTM $M$
are the elements of $\ell_1^2(\GConf_M)$ (namely, the unit vectors of
$\ell^2(\GConf_M)$). Since there is no bound on the
size of the tape in a configuration, the set $\GConf_M$ is infinite
and the Hilbert space of the configurations $\ell^2(\GConf_M)$ is
infinite dimensional.

\begin{Definition}[quantum configurations]\label{Def:quantum-config}
  Let $M$ be a pQTM.  The elements of
  $\ell^2_1(\GConf_M)$ %
  are the \textit{quantum configurations} (or more succinctly, \textit{q-configurations}) of $M$.
\end{Definition} 

We shall use Dirac notation (see Appendix~\ref{sec:HS}) for the
elements $\phi,\psi$ of $\ell^2(\GConf_M)$, writing them
$\ket{\phi}, \ket{\psi}$.

\begin{Definition}[computational basis]
  For any set of configurations $\B\subseteq \CConf_{M}$ and any
  $C\in\B$, let $\ket{C}:\B \to \CC$ be the function
  \[
  \ket{C}(D)= \left\{
    \begin{array}{ll}
      1\;\;&\mbox{if}\;C=D\\
      0\;\;&\mbox{if}\;C\neq D.
    \end{array}
  \right. 
  \]
  The set $\cb{\B}$ of all such functions is a Hilbert basis
  for $\ell^2(\B)$ (see, e.g., \cite{NiOz02}). In particular,
  following the literature on quantum computing, $\cb{\GConf_M}$ is
  called the \emph{computational basis} of $\ell^2(\GConf_M)$.  Each
  element of the computational basis is called \emph{base
    q-configuration}.
\end{Definition}

With a little abuse of language, we shall write $\ket{C}\in\ket{\phi}$
when $\phi(C)\neq 0$.  The \emph{span of} $\cb{\B}$, denoted by
$\SPAN{\cb{\B}}$, is the set of the finite linear combinations with
complex coefficients of elements of $\cb{\B}$; $\SPAN{\B}$ is a vector
space, but not a Hilbert space. In order to get a Hilbert space from
$\SPAN{\cb{\B}}$ we have to complete it, and $\ell^2(\B)$ is indeed
the unique (up to isomorphism) \emph{completion} of
$\SPAN{\cb{\B}}$~(see \cite{BerVa97}).  As a consequence, any
bounded linear
operator $U$ on $\SPAN{\cb{\B}}$ has a unique extension on
$\ell^2(\B)$, and, when $U$ is unitary, its extension is unitary
too.

For some basic definitions, properties and notations on Hilbert spaces
with denumerable basis, see Appendix~\ref{sec:HS}. In particular,
subsection~\ref{ssec:dirac-notation} presents a synoptic table of the
so-called Dirac notation that we shall use in the paper.

\subsection{Time evolution operator and QTM}
\label{ssec:time-ev-op}





In order to define the evolution operator of a pQTM $M$, it suffices
to give its behaviour on the computational basis $\cb{\GConf_M}$. In
particular, we have to distinguish three cases:
\begin{enumerate}
\item $C \in \SConf{0}_M\setminus\SConf{t}_M$. \\
  Let $C_{p,v,d}\in \SConf{0}_M\setminus\SConf{s}_M$ be the
  configuration obtained by leaving the counter to $0$, by replacing
  the symbol $u$ in the current cell with the symbol $v$, by moving
  the head on the $d$ direction, and by setting the machine into the
  new state $p$. In detail, if 
$C \simeq \langle \alpha w, q, u\beta,i,0\rangle$

we have
  \[
    C_{p,v,d} \simeq
    \begin{cases}
      \langle \alpha wv, p, \beta, i+1, 0 \rangle &\qquad
      \mbox{when $d = R$} \\
      \langle \alpha, p, wv\beta, i-1, 0 \rangle &\qquad
      \mbox{when $d = L$.}
    \end{cases}
  \]
  and we define
  \[
    W_{0,M}(\ket{C}) =
    \sum_{(p,v,d)\in (\Q\setminus \Q_s)\times\Sigma\times\DD}
    \delta_0(q,u)(p,v,d)\,\ket{C_{p,v,d}}
  \]
  where $\delta_0$ is the quantum transition function of $M$, and, as already introduced at the end of Section~\ref{ssec:transitions-qtm}, 
  $C_{p,v,d}$ is is the new
configuration obtained from $C$ by changing the current state from $q$
to $p$, by replacing the current symbol $u$ with $v$, and by moving
the tape head in the direction $d$.

\item $C \in \SConf{s}_M\setminus \SConf{0}_M$. \\
  Let
  $C_{-1}\in \SConf{s}_M$ be the source configuration obtained by
  decreasing by $1$ the counter of $C$, we define
  \[
    W_{s,M}(\ket{C}) = \ket{C_{-1}}
  \]
\item $C \in \SConf{t}_M$. \\
  Let $C_{+1}\in \SConf{t}_M\setminus \SConf{0}_M$ be the target
  configuration obtained by increasing by $1$ the counter of $C$, we
  define
  \[
    W_{t,M}(\ket{C}) = \ket{C_{+1}}
  \]
\end{enumerate}

We have then three linear operators
\begin{align*}
  W_{0,M} & : \SPAN{\cb{\SConf{0}_M \setminus \SConf{t}_M}} \to 
            \SPAN{\cb{\SConf{0}_M \setminus \SConf{s}_M}} \\
  W_{s,M} & : \SPAN{\cb{\SConf{s}_M \setminus \SConf{0}_M}} \to 
            \SPAN{\cb{\SConf{s}_M}} \\
  W_{t,M} & : \SPAN{\cb{\SConf{t}_M}} \to 
            \SPAN{\cb{\SConf{t}_M \setminus \SConf{0}_M}} %
\end{align*}  
defined on three disjoint subspaces. 
Moreover, since even the images of these three operators are disjoint,
their sum
\[
  W_M = W_{0,M} + W_{s,M} + W_{t,M}
\]
defines an automorphism on the linear space of q-configurations
$$W_M:\SPAN{\cb{\GConf}}\to\SPAN{\cb{\GConf}}.$$
Indeed, since $W_M$ is bounded, it extends in a unique
way to a continuous operator on the Hilbert space of q-configurations.

\begin{Definition}[time evolution operator]\label{Def:TimeEvolutionOp}
  The \emph{time evolution operator} of $M$ is the unique continuous extension
  $$U_M:\ell^2(\GConf_M)\to \ell^2(\GConf_M)$$
  of the bounded linear operator
  $W_M:\SPAN{\cb{\GConf}}\to\SPAN{\cb{\GConf}}$.
\end{Definition}

As it is the case for the operator $W_M$, also the time evolution operator $U_M$ can be
decomposed into three operators
\begin{align*}
  U_{0,M} & : \ell^2(\SConf{0}_M \setminus \SConf{t}_M)\to 
            \ell^2(\SConf{0}_M \setminus \SConf{s}_M) \\
  U_{s,M} & : \ell^2(\SConf{s}_M \setminus \SConf{0}_M) \to 
            \ell^2(\SConf{s}_M)\\
  U_{t,M} & : \ell^2(\SConf{t}_M)\to 
            \ell^2(\SConf{t}_M \setminus \SConf{0}_M) %
\end{align*}
s.t.\ 
\[
  U_M = U_{0,M} + U_{s,M} + U_{t,M}.
\]

\begin{Definition}[QTM]
  \label{def:qtm}
  A pQTM is a Quantum Turing Machine (QTM) when its time evolution
  operator $U_M$ is unitary.
\end{Definition}

\noindent\change{As usual, $U^{i}_M$ (the $i$-th iterate of $U_M$) is defined as $U^0_M = id$, and $U^{i+1}_M = U_M \circ U^{i}_M$.}


\subsection{Computations}
\label{ssec:computations}

A computation of a QTM is an iteration of its evolution operator on
some q-configuration. Since the time evolution operator of a QTM is
unitary, it preserves the norm of its argument and maps
q-configurations into q-configurations.  By the way, this holds for
the inverse $U_M^{-1}$ of the time evolution operator, too.

\begin{Property}
  Let $M$ be a QTM. If $\ket{\phi}\in\ell^2_1(\GConf_M)$, then
  $U_M^i\ket{\phi}\in\ell^2_1(\GConf_M)$, for every $i\in\ZZ$.
\end{Property}

\begin{Definition}[initial and final configurations]
  A q-configuration $\ket{\phi}$ is \emph{initial} when
  $\ket{\phi}\in\ell^2_1(\IConf_M)$ and is \emph{final} when
  $\ket{\phi}\in\ell^2_1(\FConf_M)$. %
  By $\ket{\nstring{n}}$ we denote the initial configuration
  $\langle\lambda,q_0,\nstring{n}, 0, 0\rangle \in \IConf_M$.
\end{Definition}

\begin{Definition}[computations]
  Let $M$ be a QTM and let $U_M$ be its time evolution operator.  For
  an initial q-configuration $\ket{\phi}\in\ell^2_1(\IConf)$, the
  \emph{computation} of $M$ on $\ket{\phi}$ is the denumerable
  sequence $\{\ket{\phi_i}\}_{i\in \NN}$ s.t.
  \begin{enumerate}
  \item $\ket{\phi_0}=\ket{\phi}$;
  \item $\ket{\phi_{i}}=U_M^i\ket{\phi}$.
  \end{enumerate}
\end{Definition}

Clearly, any computation of a QTM $M$ is uniquely determined by its
initial q-configuration. The computation of $M$ on the initial
q-configuration $\ket{\phi}$ will be denoted by $K_{\ket{\phi}}^M$.

\begin{Remark}\label{rem:final-stable}
  The definition of the time evolution operator ensures that the final
  configurations reached along a computation are stable and do not
  interfere with other branches of the computation in superposition,
  which may enter into a final configuration later. Indeed, given a
  configuration $\ket{\phi} = \ket{\phi_f} +\ket{\phi_{nf}}$, where
  $\ket{\phi_f}\in\ell^2(\FConf)$ and $\ket{\phi_{nf}}$ does not contain any
  final configuration, let
  $\ket{\psi}=U^i\ket{\phi}=U^i \ket{\phi_f} + U^i\ket{\phi_{nf}}$.  Any
  final configuration in $U^i\ket{\phi_{nf}}$ has a value of the
  counter less than $i$, while any final configuration
  $\ket{C, k} \in \ket{\phi}$ formed of a plain configuration $C$ and
  a counter $k$ is mapped into a configuration
  $\ket{C,i+k}\in\psi$. Moreover, $\ket{C,k}$ and $\ket{C,i+k}$ have
  the same coefficient in $\ket{\phi}$ and $\ket{\psi}$, respectively,
  since
  $\kinprod{\psi}{C, i+k}=\inprod{U^i\ket{\phi}}{U^i\ket{C,k}}=
  \kinprod{\phi}{C,k}$.
\end{Remark}

%

\subsection{Local conditions for unitary evolution}
\label{ssec:local-cond-unit}

In analogy to the main approaches in the literature
\cite{BerVa97,NishOza10}, it is possible to state a set of local
condition for the quantum transition function $\delta_0$ of pQTMs in
order to ensure that the time evolution operator is unitary.

\begin{Theorem}\label{theor:local}
  Let $M$ be a pQTM with quantum transition function $\delta_0$.  The
  \emph{time evolution operator}
  $U_M:\ell^2(\GConf_M)\to \ell^2(\GConf_M)$ of $M$ is unitary iff
  $\delta_0$ satisfies the local conditions:
  \begin{enumerate}
  \item for any $(q,a)\in (\Q\setminus \Q_t) \times \Sigma$
    \[
      \sum_{(p,b,d)\in (\Q \setminus \Q_s)\times\Sigma\times\DD}
      |\delta_0(q,a)(p,b,d)|^2=1
    \]
  \item for any $(q,a), (q',a')\in (\Q \setminus \Q_t)\times\Sigma$ with $(q,a)\neq(q',a')$
    \[
      \sum_{(p,b,d)\in (\Q \setminus \Q_s)\times\Sigma\times\DD} 
      \delta_0(q',a')(p,b,d)^*\delta_0(q,a)(p,b,d)=0
    \]
  \item for any $(q,a,b), (q',a',b')\in (\Q \setminus \Q_t)\times\Sigma^2$
    \[
      \sum_{p\in (\Q \setminus \Q_s)} 
      \delta_0(q',a')(p,b',L)^*\delta_0(q,a)(p,b,R)=0
    \]
  \end{enumerate}
    where $()^*$ denote the complex conjugate operator.
\end{Theorem}
\begin{proof}
  We show that, to any QTM $M$ (as defined above), we can associate a
  \BeV QTM $N$ which differs on the behaviour on source and target
  states only. The unitarity of $M$ follows then from the unitarity of $N$.
  
  Let $M=\langle \Sigma, \Q, \Q_s, \Q_t, \delta_0, q_i, q_f\rangle$.
  We take the \BeV QTM $N$ with the same states $\Q$ and the same
  alphabet $\Sigma$, and whose transition function $\rho$ satisfies
  the following conditions:
  \begin{enumerate}
  \item $\rho$ is the same of the transition function $\delta_0$ of
    $M$ for any non-terminal state;
  \item for any terminal state $q_t$ and every current symbol $u$,
    $\rho$ has a unique non-null transition, with weight $1$, that
    leaves the current symbol on the tape unchanged, moves the head to
    the right, and goes into an initial state $q_s$;
  \item the non-null out transitions from two distinct terminal states
    lead to two distinct source states, defining in this way a
    bijection between the set of the target and source states.
  \end{enumerate}

  In other words, $\rho$ is defined by:
  \begin{align*}
    \rho(q,u)(p,v,d)&= \delta_0(q,u)(p,v,d) &&\qquad \mbox{for $q\in\Q\setminus\Q_t$,
                                               $p\in\Q\setminus\Q_s$, $u,v\in\Sigma$, $d\in\DD$} \\
    \rho(q,u)(s,v,d)&= 0 &&\qquad \mbox{for $q\in\Q\setminus\Q_t$,
                            $s\in\Q_s$, $u,v\in\Sigma$, $d\in\DD$}
                            \intertext{and, assuming $\Q_s=\{s_i\}_{0\leq i \leq k}$ and $\Q_t=\{t_i\}_{0\leq i \leq k}$}
                            \rho(t_i,u)(s_i,u,R)&= 1 &&\qquad \mbox{for $u\in\Sigma$} \\
    \rho(t_i,u)(s_i,v,L)&= 0 &&\qquad \mbox{for $u,v\in\Sigma$} \\
    \rho(t_i,u)(q,v,d)&= 0 &&\qquad \mbox{for $q\in\Q\setminus\Q_t$, $q\neq s_i$,
                              $u,v\in\Sigma$, $d\in\DD$}
  \end{align*}

  The above definition ensures that $N$ is a \BeV QTM whose
  configurations are the plain configurations of $M$, and s.t.\ its
  transition function $\rho$ verifies \BeV local conditions.
  By \BeV, the evolution operator $U_N$ of $N$ is then a unitary
  operator on the Hilbert space of the plain configurations of $N$
  and, as a consequence, its adjoint $U_N^*$ is its inverse.
  We can take three linear operators

  \begin{align*}
    V_{0,M} & : \SPAN{\cb{\SConf{0}_M \setminus \SConf{s}_M}}  \to
              \SPAN{\cb{\SConf{0}_M \setminus \SConf{t}_M}}
              \\
    V_{s,M} & : \SPAN{\cb{\SConf{s}_M}} \to
              \SPAN{\cb{\SConf{s}_M \setminus \SConf{0}_M}}
              \\
    V_{t,M} & : \SPAN{\cb{\SConf{t}_M \setminus \SConf{0}_M}} \to
              \SPAN{\cb{\SConf{t}_M}}
  \end{align*}  
  defined by
  \begin{align*}
    V_{0,M} \ket{C} &= \ket{D}
    && \mbox{with $\ket{D} = U_N^* \ket{C}$
       when $\ket{C} \in \SConf{0}_M\setminus\SConf{s}_M$}
    \\
    V_{s,M} \ket{C} &= \ket{C_{+1}}
    &&
       \mbox{when $\ket{C} \in \SConf{s}_M$}
    \\
    V_{t,M} \ket{C} &= \ket{C_{-1}}
    &&
       \mbox{when $\ket{C} \in \SConf{t}_M\setminus\SConf{0}_M$}
  \end{align*}
  Which are the inverses of the three corresponding linear operators
  $W_{0,M}$, $W_{s,M}$, and $W_{tM}$, respectively (see
  subsection~\ref{ssec:time-ev-op}). Thus, the bounded linear operator
  $V_M=V_{0,M}+V_{s,M}+V_{tM}$ is the inverse of
  $W_M=W_{0,M}+W_{s,M}+W_{tM}$, and its continuous extension (obtained
  by completion) is the inverse of $U_M$. Moreover, such an operator
  is the adjoint $U_M^*$ of $U_M$ also. Summing up, the adjoint $U_M^*$ of
  $U_M$ is its inverse and, as a consequence, $U_M$ is unitary.
\end{proof}

\begin{Remark}
  The reader may wonder why we need to have cell indexes in
  configurations, when \change{they are not used in many definitions of} classical TMs. In
  fact, cell indexes are needed only in a very special case, that of a
  completely empty tape.  Let us suppose, indeed, to eliminate
  indexes, and to feed a completely blank tape to a QTM with only two
  states $\{q,r \}$, with no source and final states and no tape
  symbols except $\Box$; let the transition function $\delta_0$ be
  s.t.
  \begin{itemize}
  \item $\delta_0(q,\Box)(q,\Box,L)= 1/2$
  \item $\delta_0(q,\Box)(q,\Box,R)= 1/2$
  \item $\delta_0(q,\Box)(r,\Box,L)= 1/2$
  \item $\delta_0(q,\Box)(r,\Box,R)= -1/2$
  \item $\delta_0(r,\Box)(q,\Box,L)= 1/2$
  \item $\delta_0(r,\Box)(q,\Box,R)= -1/2$
  \item $\delta_0(r,\Box)(r,\Box,L)= 1/2$
  \item $\delta_0(r,\Box)(r,\Box,R)= 1/2$
  \end{itemize}
  It is easy to check that the conditions of Theorem~\ref{theor:local}
  are satisfied, but trivially the induced time evolution operator $U$
  is not an isometry since, given the empty string $\lambda$ we have
  $U(\lambda,q,\lambda)=
  U(\lambda,r,\lambda)=(\lambda,q,\lambda)$.\footnote{We omit here
    the counter, which always remains set to zero.}  The indexing we
  use is consistent with the one adopted by \Bev,
  \cite{BerVa97}--Definition 3.2, pag.~1419.  Here we just want to
  point out that \Bev{} only provides a fairly light informal definition
  of configuration. We have preferred a more pedantic formulation,
  which in our opinion is necessary to eliminate all the reader's
  doubts about the correctness of the proposed theorems.

\end{Remark}

\subsection{A comparison with Bernstein and Vazirani's QTMs: part 1}
\label{ssec:comp-BandV-1}

We refer to \BeV~\cite{BerVa97} for the precise definitions of the QTMs used in that
paper. For the sake of readability, we informally recall the notion of
what they call \textit{well formed, stationary, normal form QTMs}
(\BeV-QTMs in the following).


A \BeV-QTM $M=\langle \Sigma, \Q, \delta, q_0, q_f\rangle$ is defined
as our QTM (with one source state and one target state only) with the
following differences:
\begin{enumerate}
\item the set of configurations coincides with all possible classical
  configurations, namely all the set
  $\Sigma^*\times \Q\times\Sigma^*\times\ZZ$;
\item no superposition is allowed in the initial q-configuration (it
  must be a classical configuration $\langle\alpha,q,\beta,i\rangle$
  with amplitude 1);
\item given an initial configuration $\ket{C}$, the q-configuration
  $\phi_k=U_{M}^{k}\ket{C}$ is the \emph{final q-configuration} for
  the computation starting in $\ket{C}$, when: (i) all the
  configurations in $\phi_k$ are final; (ii) for all $i < k$, the
  q-configuration $\phi_i={U_M}^i\ket{C}$ does not contain any final
  configuration. When this is the case, we also say that the QTM
  \emph{halts in} $k$ steps in $\phi_k$;
\item if a QTM halts, then the tape head is on the start cell of the
  initial configuration;
\item there is no counter and for every symbol $a$ there is loop from
  the only final state $q_f$ into the only initial state $q_0$, that
  is, $\delta(q_f,a)(q_0,a,R) = 1$ for every $a\in\Sigma$. Therefore,
  because of the local unitary conditions (that must hold in the final
  state too), these are the only outgoing transitions from $q_f$, and
  the only incoming ones into $q_0$.  So, $\delta(q_f,a)(q',a',d) = 0$
  if $(q',a',d)\neq(q_0,a,R)$ and $\delta(q',a')(q_0,a,d)=0$ if
  $(q',a',d) \neq (q_f,a,R)$.
\end{enumerate}

\begin{Theorem}\label{theor:fromBeVtoOur}
  For any \BeV-QTM $M$ there is a  QTM $M'$ s.t.\ for each initial configuration
  $\ket{C}$, if $M$ with input $\ket{C}$ halts in $k$ steps in a final configuration
  $\ket{\phi}=U_M^k\ket{C}$, then $U_{M'}^k{\ket{C}}=\ket{\phi}$.
\end{Theorem}
\begin{proof}
  The QTM $M'$ has the same states of $M$, the initial state $q_0$ is
  its only source state, and the final state $q_f$ is its only target
  state. Therefore, if
  $M=\langle \Sigma, \Q, \delta, q_0, q_f\rangle$, we take
  $M'=\langle \Sigma, \Q, \{q_0\}, \{q_f\}, \delta_0, q_0,
  q_f\rangle$, where $\delta_0$ is the restriction of $\delta$ to
  $((\Q\setminus \Q_t)\times\Sigma) \to \ell^2((\Q\setminus \Q_s)
  \times \Sigma \times \DD)$, that is, for each $q\neq q_f$ and
  $a\in\Sigma$, we have $\delta_0(q,a)(p,b,d)=\delta(q,a)(p,b,d)$, for
  every $(p,b,d)\in (\Q\setminus \Q_s) \times \Sigma\times \DD$.
  Since the local unitary conditions hold for $\delta$ and, in a
  \BeV-QTM, $\delta(q,a)(q_0,b,d)=0$, when $q\neq q_f$, the unitary
  local conditions hold for $\delta_0$ too.
  
  By construction, if $U_{M}^i{\ket{C}}$ is not final for
  $0 \leq i< k$, then
  $\ket{\phi_k}=U_{M}^k{\ket{C}}=U_{M'}^k{\ket{C}}$. In particular,
  this holds when $\ket{\phi_k}$ is the final configuration of the
  \BeV-QTM $M$.
\end{proof}

  We remark that the inverse cannot hold true: 
A \BeV-QTM for a given input cannot exhibit both a
  non terminated computation and a terminated one (see
  Example~\ref{ssec:ex:identity} for a machine that cannot be simulated by
  a \BeV-QTM{}).

\section{Quantum Computable Functions}
\label{sec:quant-comp-fun}

In this section we address the problem of defining the concept of
quantum computable function in an ``ideal'' way, without taking into
account any measurement protocol. The problem of the observation
protocol will be addressed in Section~\ref{sec:observables}.  Here we
show how each QTM naturally defines a computable function from the
sphere of radius $1$ in $\ell^2(\GConf_M)$ to the set of (partial)
probability distributions on the set of natural numbers.

\subsection{Probability distributions}
\label{ssec:prob-dist}

\begin{Definition}[Probability distributions]
  \mbox{}
  \begin{enumerate}
  \item A \emph{partial probability distribution} (PPD) of natural
    numbers is a function $\pd{}: \NN \to \mathbb{R}_{[0,1]}$ such
    that $\sum_{n\in\NN} \pd{}(n)\leq 1$.
  \item  If $\sum_{n\in\NN} \pd{}(n)= 1$, $\pd{}$ is a \emph{probability
      distribution} (PD).
  \item $\PP$ and $\PP_1$ denote the sets of all
    the PPDs and PDs, respectively.
  \item If the set $\{n : \pd{}(n)\neq 0\}$ is finite, $\pd{}$ is
    \emph{finite}.
  \item Let $\pd{}', \pd{}''$ be two PPDs, we say that
    $\pd{}' \leq \pd{}''$ ($\pd{}' < \pd{}''$) iff for each
    $n\in \NN$, $\pd{}'(n)\leq \pd{}''(n)$ ($\pd{}'(n) < \pd{}''(n)$).
  \item Let $\varP=\{\pd{i}\}_{i\in\NN}$ be a denumerable
    sequence of PPDs; $\varP$ is \emph{monotone} iff
    $\pd{i}\leq \pd{j}$, for each $i<j$.
  \end{enumerate}
\end{Definition}

\begin{Remark}
\label{rem:ppds}
  In the following, we shall also use the notation
  $\pd{}(\bot)=1-\sum_{n\in\NN}\pd{}(n)$. By definition,
  $0\leq \pd{}(\bot) \leq 1$, and a PPD is a PD iff $\pd{}(\bot)=0$.
  We also stress that $\leq$ is a partial order of PPDs and that any
  PD $\pd{}$ is maximal w.r.t.\ to $\leq$.
\end{Remark}


\begin{Definition}[limit of a sequence of PPDs]
  Let $\varP=\{\pd{i}\}_{i\in\NN}$ be a sequence of PPDs.  If for each
  $n\in\NN$ there exists $l_n=\lim_{i\to \infty} \pd{i}(n)$, we say
  that $\lim_{i\to \infty} \pd{i}=\pd{}: \NN \to \mathbb{R}_{[0,1]}$,
  with $\pd{}(n)=l_n$.
\end{Definition}

\subsection{Monotone sequences of probability distributions}
\label{ssec:mon-seq-pd}

\begin{Proposition}
  \label{prop:mon-seq}
  Let $\varP=\{\pd{i}\}_{i\in\NN}\subseteq\PP$ be a monotone sequence
  of PPDs.
  \begin{enumerate}
  \item\label{item:mon-seq:sup-lim}
    $\lim_{i\to\infty}\pd{i}$ exists and it is the supremum $\lub \varP$
    of $\varP$ ;
  \item\label{item:mon-seq:comm-lim}
    $\sum_{n\in\NN} (\lub \varP) (n) = \lub \left\{\sum_{n\in\NN}
      \pd{i}(n)\right\}_{i\in\NN}$;
  \item\label{item:mon-seq:sup-ppd} $\lub \varP \in \PP$.
  \end{enumerate}
\end{Proposition}
\begin{proof}
  Since $\pd{i}(n) \leq 1$, every non-decreasing sequence
  $\{\pd{i}(n)\}_{i\in\NN}$ has a supremum
  $\sup\{\pd{i}(n)\}_{i\in\NN}=\lim_{i\to\infty}\pd{i}(n)$. Thus,
  $\lim_{i\to\infty}\pd{i}$ is defined and
  $\pd{i} \leq \lim_{i\to\infty}\pd{i}$, for every $i\in\NN$.  On the
  other hand, for any $\pd{}'$ s.t.\ $\pd{i} \leq \pd{}'$ for
  $i\in\NN$, we have
  $\lim_{i\to\infty}\pd{i}(n) = \sup \{\pd{i}(n)\}_{i\in\NN}\leq
  \pd{}'(n)$, for every $n\in\NN$; namely,
  $\lim_{i\to\infty}\pd{i}\leq\pd{}'$. We can then conclude
  (item~\ref{item:mon-seq:sup-lim}) that
  $\lub \varP =\lim_{i\to\infty}\pd{i}$.

  Let us now prove item~\ref{item:mon-seq:comm-lim}. First of all,
  since $0 \leq \pd{i}(n) \leq (\lub \varP) (n)$ for $i,n\in \NN$, we
  have
  $\sum_{n\in\NN} (\lub \varP) (n) \geq \sup \left\{\sum_{n\in\NN}
    \pd{i}(n)\right\}_{i\in\NN}$ and
  \begin{multline*}
    \qquad\qquad %
    \sup \left\{\sum_{n\in\NN}\pd{i}(n)\right\}_{i\in\NN} %
    \geq \sup\left\{\sum_{n\leq k}\pd{i}(n)\right\}_{i\in\NN} %
    \\
    = \sum_{n\leq k}\sup\left\{\pd{i}(n)\right\}_{i\in\NN} %
    =\sum_{n\leq k} (\lub\varP)(n) %
    \qquad\qquad %
  \end{multline*}
  for any $k\in\NN$. Thus, 
  \begin{multline*}
    \qquad\qquad %
    \sup \left\{\sum_{n\in\NN}\pd{i}(n)\right\}_{i\in\NN} %
    \geq \sup\left\{\sum_{n\leq k} (\lub\varP)(n)\right\}_{k\in\NN} %
    \\
    = \lim_{k \to \infty}\sum_{n\leq k} (\lub\varP)(n) %
    = \sum_{n\in\NN} (\lub\varP)(n) %
    \qquad\qquad %
  \end{multline*}
  
  Summing up, we can conclude, as
  \[
    \sum_{n\in\NN} (\lub \varP) (n) \geq \sup \left\{\sum_{n\in\NN}
    \pd{i}(n)\right\}_{i\in\NN} \geq \sum_{n\in\NN} (\lub\varP)(n)
  \]
  
  Finally, by hypothesis, $\sum_{n\in\NN}\pd{i}(n) \leq 1$, for any
  $i\in\NN$. Therefore,
  $\sum_{n\in\NN} (\lub \varP) (n) = \sup \left\{\sum_{n\in\NN}
    \pd{i}(n)\right\}_{i\in\NN} \leq 1$. Which proves
  (item~\ref{item:mon-seq:sup-ppd}) that $\lub \varP \in \PP$.
\end{proof}

\subsection{PPD sequence of a computation}
\label{sec:ppd-seq-comp}

The computed output of a QTM will be defined
(Definition~\ref{def:comp-out}) as the limit of the sequence of
PPDs obtained along its computations.

\begin{Definition}[probability and q-configurations]
  \label{def:prob-comf}
  Given a configuration $C=\langle\alpha,q,\beta,i,n\rangle$, let
  $\val[C]$ be the number of $1$s in $\alpha\beta$.  For any
  $\ket{\phi}\in\ell^2(\CConf)$, let us define
  $\pd{\ket{\phi}}: \NN \to \mathbb{R}_{\geq 0}$ s.t.\
   \[
     \prob{\ket{\phi}}{n}=\sum_{C\in\FConf, \val[C] =n} |e_C|^2
   \]
   when $\ket{\phi}=\sum_{C\in\CConf} e_C \ket{C}$.
\end{Definition}

\begin{Proposition}
  \label{prop:ppd-qconf}
  If $\ket{\phi}$ is a q-configuration, $\pd{\ket{\phi}}$ is a
  PPD. Moreover, it is a PD iff $\ket{\phi}$ is final.
\end{Proposition}
\begin{proof}
  Let $\ket{\phi} = \ket{\phi_0} + \ket{\phi_f}$ with
  $\ket{\phi_0}\in\ell^2(\CConf_M\setminus\FConf_M)$, and
  $\ket{\phi_f}\in\ell^2(\FConf_M)$. By the definition of
  $\pd{\ket{\phi}}$, it is readily seen that
  $\pd{\ket{\phi}}=\pd{\ket{\phi_f}}$ and that
  $\sum_{n\in\NN}\pd{\ket{\phi}}(n)=
  \sum_{n\in\NN}\pd{\ket{\phi_f}}(n)= \norm{\ket{\phi_f}}\leq
  1$. Therefore, $\pd{\ket{\phi}}$ is a PPD. Moreover, since
  $\norm{\ket{\phi_f}}=1$ iff $\norm{\ket{\phi_0}}=0$.  We see that
  $\pd{\ket{\phi}}$ is a PD iff $\ket{\phi_0}=0$, that is, iff
  $\ket{\phi}=\ket{\phi_f}$.
\end{proof}

\begin{Definition}
  \label{def:prob-qconf}
  For any q-configuration $\ket{\phi}$, we shall 
   say that $\pd{\ket{\phi}}$ is the PPD associated to $\ket{\phi}$,
   and we shall denote by $\pd{K_{\ket{\phi}}^M}$ the sequence of PPDs
   $\{\pd{\ket{\phi_i}}\}_{i\in \NN}$ associated to the computation
   $K_{\ket{\phi}}^M=\{\ket{\phi_i}\}_{i\in \NN}$.
\end{Definition}

The PPD sequence of any QTM computation is monotone. In the simple
proof of this key property (Theorem~\ref{theor:monot}) we see at work
all the constraints on the time evolution of a QTM $M$.
\begin{enumerate}
\item That when in a final (target) configuration, the machine can only
  increment the counter; as a consequence, the $\val$ of final (target) 
  configurations does not change.
\item That when entering for the first time into a final (target)
  state, the value of the counter is initialised to $0$.
\item That when in a final (target) configuration $\ket{\phi, n}$,
  the counter gives the number of steps $n$ since $M$ is looping into
  the plain configuration $\phi$.
\end{enumerate}
We stress that the last two properties defuse quantum interference between final
configurations reached in a different number of steps.

\begin{Theorem}[monotonicity of computations]\label{theor:monot}
  For any computation $K_{\ket{\phi}}^M$ of a QTM $M$, the sequence of
  PPDs $\pd{K_{\ket{\phi}}^M}$ is monotone.
\end{Theorem}
\begin{proof}
  Let us prove that $\pd{\ket{\phi}} \leq \pd{U\ket{\phi}}{}$,
  for every $\ket{\phi}\in\ell^2(\CConf_M)$. Let
  $\ket{\phi} = \ket{\phi_0} + \sum_{k\in\NN} \ket{\phi_{f,k}}$, with
  $\ket{\phi_0}\in\ell^2(\CConf_M\setminus\FConf_M)$, and
  $\ket{\phi_{f,k}}\in\ell^2(\FConf_M\cap\SConf{k}_M)$, where
  $\SConf{k}_M$ is the set of the configurations of $M$ whose counter
  is equal to $k\in\NN$.  For every $n\in\NN$, we see that
  $\prob{\ket{\phi}}{n}=\sum_{k\in\NN}\prob{\ket{\phi_{f,k}}}{n}$. Let
  $\ket{\psi} = \ket{\psi_0} + \sum_{k\in\NN}
  \ket{\psi_{f,k}}=U\ket{\phi}$. By the definition of $U$, we see that
  $U\ket{\phi_0}= \ket{\psi_0}+\ket{\psi_{f,0}}$ and $U\ket{\phi_{f,k}}=\ket{\psi_{f,k+1}}$.
  Therefore, $\prob{\ket{\psi}}{n}=\sum_{k\in\NN}\prob{\ket{\psi_{f,k}}}{n}=
  \prob{\ket{\psi_{f,0}}}{n}+\sum_{k\in\NN}\prob{\ket{\phi_{f,k}}}{n}=
  \prob{\ket{\psi_{f,0}}}{n}+\prob{\ket{\phi}}{n}\geq \prob{\ket{\phi}}{n}$.
\end{proof}

\subsection{Computed output}
\label{ssec:computed-output}

We can now come back to the definition of the computed output of a QTM
computation.  The easy case is when a computation reaches a final
q-configuration $\ket{\psi}\in \FConf$ (meaning that all the classical
computations in superposition are ``terminated'')---in this case the
computed output is the PD $\pd{\ket{\psi}}$. The QTM keeps computing
and transforming $\ket{\psi}$ into other configurations, but all these
configurations have the same PD.  However, we want to give meaning
also to ``infinite'' computations, which never reach a final
q-configuration, yet producing some final configurations in the
superpositions.  For this purpose, we define the computed output as
the limit of the PPDs yielded by the computation. 




\begin{Definition}[computed output of a QTM]
  \label{def:comp-out}
  Let $K_{\ket{\phi}}^M=\{\ket{\phi_i}\}_{i\in\NN}$ be the computation
  of the QTM $M$ on the initial q-configuration $\ket{\phi}$.
%
%
  The computed output of $M$ on the initial q-configuration
  $\ket{\phi}$ is the PPD $\pd{}=\lim_{i\to \infty}\pd{\ket{\phi_i}}$,
  which we shall also denote by $\lim K^M_{\ket{\phi}}$, or by the
  notation $M_{\ket{\phi}}\to \pd{}$.
\end{Definition}

Let us remark that, by Proposition~\ref{prop:mon-seq} and
Theorem~\ref{theor:monot}, the limit in the above definition is
well-defined for any computation. Therefore, a QTM has a computed
output for any initial q-configuration $\ket{\phi}$.

%

\begin{Definition}[finitary computations]
  Given a QTM $M$, a q-configuration $\ket{\phi}$ is \emph{finite} if
  it is an element of $\mathsf{span(CB(\GConf_M))}$.
  A computation $K_{\ket{\phi}}^M=\{\ket{\phi_i}\}_{i\in \NN}$ is
  \emph{finitary} with computed output $\pd{}$ if there exists a $k$
  s.t. $\ket{\phi_k}$ is final and $\pd{\ket{\phi_k}}=\pd{}$.
\end{Definition}

\begin{Proposition}\label{prop:partiality}
  Let $K_{\ket{\phi}}^M=\{\ket{\phi_i}\}_{i\in \NN}$ be a finitary
  computation with  computed output $\pd{}$, that is, $M_{\ket{\phi}}\to \pd{}$.
  \begin{enumerate}
  \item There exists a $k$, such that for each $j\geq k$, $\ket{\phi_j}$ is final 
  and $\pd{\ket{\phi_j}}=\pd{}$.
  \item $\pd{}$ is a PD. 
  \end{enumerate}
\end{Proposition}
\begin{proof}
  By definition, there is a $k$ s.t.\ $\ket{\phi_k}$ is final. Let
  $\pd{\ket{\phi_k}}=\pd{}$.  By Proposition~\ref{prop:ppd-qconf},
  $\pd{}$ is a PD. By monotonicity, $\pd{}\leq \pd{\ket{\phi_j}}$, for
  every $j \geq k$. Thus, since any PD is maximal for $\leq$ (see
  Remark~\ref{rem:ppds}), $\pd{}=\pd{\ket{\phi_j}}$, for every
  $j \geq k$.
\end{proof}

Given a computation $K_{\ket{\phi}}^M$, we can then distinguish the
following cases:
\begin{enumerate}
\item $K_{\ket{\phi}}^M$ is finitary. In this case
  $M_{\ket{\phi}}\to \pd{}\in\PP_1$; the output of the computation is
  then a PD and is determined after a finite number of steps;
\item $K_{\ket{\phi}}^M$ is not finitary, but
  $M_{\ket{\phi}}\to \pd{}\in\PP_1$. The output is a PD and is
  determined as a limit;
\item $K_{\ket{\phi}}^M$ is not finitary, and
  $M_{\ket{\phi}}\to \pd{}\in\PP\setminus\PP_1$ (the sum of the probabilities
  of observing natural numbers is $p<1$). Not only the result is
  determined as a limit, but we cannot extract a  PD from the
  output.
\end{enumerate} 

The first two cases above give rise to what
Definition~\ref{definition:qcf} calls a \emph{q-total} function.
Observe, however, that for an external observer, cases (2) and (3) are
in general indistinguishable, since at any finite stage of the
computation we may observe only a finite part of the computed output.

For some examples of QTMs and their computed output, see
Section~\ref{sec:comp-BandV-2}.

\subsection{Quantum partial computable functions}
\label{ssec:quant-part-comp-fun}

We want our quantum computable functions to be defined over a natural
extension of the natural numbers.
When using a QTM for computing a
function, we stipulate that initial q-configurations are
superpositions of initial classical (plain) configurations
$\ket{C_n}$, where the tape of the configuration $C_n$ encodes the
number $n$. A natural choice for this encoding corresponds to take
$C_n=\ket{\nstring{n}}=\ket{\langle\lambda,q_0,\nstring{n},0,0\rangle}$,
where $\nstring{n}$ denotes the string $1^{n+1}$. 
Such q-configurations
are naturally isomorphic to the space
$\ell^2_1(\NN)= \left\{ \phi:\NN \rightarrow\CC \mid
  \sum_{n\in\NN}|\phi(n)|^2=1\right\}$ of square summable, denumerable
sequences with unitary norm, under the bijective mapping
$\nu(\sum d_k n_k) = \sum d_k \ket{\nstring{n_k}}$. 
\change{Observe that, for output, we simply count the number of (non contiguous) $1$s on the tape,
according to Definition~\ref{def:prob-comf}.}


\begin{Definition}[partial quantum computable functions]\label{definition:qcf}
  \mbox{}
  \begin{enumerate}
  \item A function $f:\ell^2_1 \to \PP$ is \emph{partial quantum
      computable} (q-computable) if there exists a QTM $M$
    s.t. $f(\mathbf{x})=\pd{}$ iff $M_{\nu(\mathbf{x})}\to \pd{}$.
  \item A q-partial computable function $f$ is \emph{quantum total}
    (q-total) if for each $\mathbf{x}$, $f(\mathbf{x})\in\PP_1$.
  \end{enumerate}
\end{Definition}

\change{It is out of the scope of the paper to investigate in details how this class exactly relates to other classes of computable functions when the input of QTMs is restricted to classical inputs. Observe, however, that when also the output is required to be classical, and all amplitudes are computable, we expect to obtain the classical (partial) recursive functions. Indeed, any classical deterministic TM may be simulated by a reversible TM~\cite{Ben73}; moreover, \BeV  show that reversible TMs are a special case of QTMs (their Section 4; see also our Section~\ref{Section:reversibleTM}.). On the other direction, when we have classical inputs and computable amplitudes, it is easy to simulate a QTM via a classical TM.}



\section{Observables}
\label{sec:observables}

While the evolution of a closed quantum system (e.g., a QTM) is
reversible and deterministic once its evolution operator is known, a
(global) measurement of a q-configuration is an irreversible process,
which causes the collapse of the quantum state to a new
state. Technically, a measurement corresponds to a projection on a
subspace of the Hilbert space of quantum states. For the sake of
simplicity, in the case of QTMs, let us restrict to measurements
observing if a configuration belongs to the subspace described by some
set of base configurations $\B$. The effect of such a measurement is
summarised by the following:

\medskip

\begin{quote}
  \noindent\textbf{Measurement}\\
  Given a set of configurations $\B\subseteq\GConf$, a measurement
  observing if a quantum configuration
  $\ket{\phi}=\sum_{C\in\GConf} e_C \ket{C}$ belongs to the subspace
  generated by $\cb{\B}$ gives a positive answer with a probability
  $p=\sum_{C\in\B}|e_C|^2$, equal to the square of the norm of the
  projection of $\ket{\phi}$ onto $\ell^2(\B)$, causing at the same
  time a collapse of the configuration into the normalised projection
  $\sum_{C\in\B} p^{-\frac{1}{2}}e_C \ket{C}$; dually, with probability
  $1-p=\sum_{C\not\in\B}|e_C|^2$ , it gives a negative answer and a
  collapse onto the subspace $\ell^2(\GConf\setminus\B)$ orthonormal
  to $\ell^2(\B)$, that is, into the normalised configuration
  $\sum_{C\not\in\B} (1-p)^{-\frac{1}{2}}e_C \ket{C}$, 
\end{quote}
  \change{This kind of measurement is a special case of the so called projective measurement that is a consequence of the postulates of quantum mechanics.
  	For a detailed discussion  see e.g. \cite{KLM07, axiomsMZ19}.}
\medskip

Because of the \textit{irreversible} modification produced by any
measurement on the current configuration, and therefore on the rest of
the computation, we must  deal with the problem of how to read the
result of a computation. In other words, we need to establish some
protocol to observe when a QTM has eventually reached a final
configuration, and to read the corresponding  result.

\subsection{The approach of Bernstein and Vazirani}
\label{ssec:approach-output-BeV}

We already discussed how \BeV's ``sensible'' QTMs are machines where
all the computations in superposition are in some sense terminating,
and reach the final state at the same time (are ``stationary'', in
their terminology).  More precisely, Definition~3.11 of~\BeV reads: \textit{"A final configuration of a QTM is any configuration in
  [final] state. If when QTM $M$ is run with input $x$, at time $T$
  the superposition contains only final configurations, and at any
  time less than $T$ the superposition contains no final
  configuration, then $M$ halts with running time $T$ on input $x$."}

This is a good definition for a theory of computational complexity
(where the problems are classical, and the inputs of QTMs are always
classical) but it is of limited use in the semantics of quantum programming languages. 
Indeed, inputs of a \BeV-QTM \textit{must} be
classical---we cannot extend by linearity a \BeV-QTM on inputs in
$\ell_1^2$, since there is no guarantee whatsoever that on different
inputs the same QTM halts with the same running time.

\subsection{The approach of Deutsch}
\label{sec:approach-output-deutsch}

Deutsch~\cite{Deu85} assumes that QTMs are enriched with a termination
bit $T$.  At the beginning of a computation, $T$ is set to $0$, then,
the machine sets this termination bit to $1$ when it enters into a
final configuration. If we write $\ket{T=i}$ for the function that
returns $1$ when the termination bit is set to $i$, and $0$ otherwise,
a generic q-configuration of a Deutsch's QTM can be written as

\[\ket{\phi}=
  \ket{T=0}\otimes \sum_{C\not\in \FConf} e_C\ket{C} 
  +\ket{T=1}\otimes \sum_{D\in\FConf}d_D\ket{D}
\]

The observer periodically measures $T$ in a non destructive way (that
is, without modifying the rest of the state of the machine).
\begin{enumerate}
\item If the result of the measurement of $T$ gives the value $0$,
  $\ket{\phi}$ collapses (with a probability equal to
  $\sum_{C\not\in\FConf} |e_C|^2$) to the q-configuration
  \[
    \ket{\psi'} = 
    \frac{%
      \ket{T=0}\otimes\sum_{C\not\in\FConf}e_C\ket{C}}{%
      \sqrt{\sum_{C\not\in\FConf}|e_C|^2}
    }
  \]
  and the computation continues with $\ket{\psi'}$.
\item If the result of the measurement of $T$ gives the value $1$,
  $\ket{\phi}$ collapses (with probability
  $\sum_{D\in\FConf} |d_D|^2$) to
  \[
    \ket{\psi''} =
    \frac{%
      \ket{T=1}\otimes\sum_{D\in\FConf} d_D\ket{D}}{%
      \sqrt{\sum_{D\phantom{\not}\in\FConf}|d_D|^2}
    }
  \]
  and, immediately after the collapse, the observer makes a further
  measurement of the component
  $\dfrac{\sum_{D\in\FConf}d_D\ket{D}}{\sqrt{\sum_{D\in\FConf}|d_D|^2}}$
  in order to read-back a final configuration.
\end{enumerate}

Note that  Deutsch's protocol (in an irreversible way) spoils at
each step the superposition of configurations.
The main point of Deutsch's approach is that a measurement must be
performed immediately after some computation enters into a final state. In
fact, since at the following step the evolution might lead the machine
to exit the final state modifying the content of the tape, we would
not be able to measure at all this output. In other words, either the
termination bit acts as a trigger that forces a measurement each time
it is set, or we perform a measurement after each step of the
computation.

\subsection{Problems with Deutsch measurement}

Since a QTM must be physically feasible, at least ``in principle'' ,
let us analyse in detail whether the termination protocol proposed by
Deutsch~\change{\cite{Deu85}} is realizable.

As any quantum system, a QTM evolves with respect to a continuos time. This
means that the time evolution operator $U$ of a machine $M$
is the 
solution of the Schrödinger equation with a given Hamiltonian $H$ and  depends
on the time~\change{\cite{Ish95}}. In particular, when $H$ is constant, the solution of
the equation is
\[ U_t = e^{-(i / h) Ht)} \quad\mbox{\ for\ }t\in \mathbb{R}^+ \]

Now, let $\Delta$ be the duration of a single step of a Deutsch QTM.
This means that after $n$ steps of the QTM, we are at time
$t = n\Delta$, and that for the corresponding configurations $\ket{\psi_{n}}$ of the QTM, it holds the following property
\begin{multline*}
  \ket{\psi_{n+1}} =  U_{(n+1)\Delta} \ket{\psi_0}  \\
  = e^{-(i / h) H (n+1)\Delta)}\ket{\psi_0}
  = e^{-(i / h) H \Delta)}e^{-(i / h) H n\Delta)}\ket{\psi_0}
  = U_\Delta U_{n\Delta}\ket{\psi_0} \\
  = U_\Delta \ket{\psi_n} 
\end{multline*}
In other words, at each step $n$, the next configuration
$\ket{\psi_{n+1}}$ is obtained by applying the evolution operator
$U_\Delta$ to the current configuration $\ket{\psi_{n}}$.

Now, let us suppose that, as a consequence of the measurement of the
halt qubit $T$, after $n$ steps, and therefore at time
$t_n=n\Delta$, the system collapses to the quantum configuration
\[
  \ket{\psi_{n}} =
  \frac{%
    \ket{T=1}\otimes\sum_{D\in\FConf} d_D\ket{D}}{%
    \sqrt{\sum_{D\phantom{\not}\in\FConf}|d_D|^2}
  }
\]
According to Deutsch's approach, one has to perform a second
measurement to read the actual value of the output on the tape. But,
since we are in a continuous time setting, what is the time of
this second measurement? Since it can be done only \textit{after} the
measure of the halt qubit $T$, this must take place at a time
$t_{n+\epsilon}=(n+\epsilon)\Delta$, with $\epsilon > 0$.
We can distinguish two cases (analogous reasonings apply for $\epsilon > 1$):
\begin{enumerate}
\item $0< \epsilon < 1$. In this case, the tape is measured at an
  intermediate time between $t_n$ and $t_{n+1}$, and the machine is in
  an intermediate configuration $\ket{\psi_{n+\epsilon}}$ between $\ket{\psi_n}$ and
  $\ket{\psi_{n+1}}$. Since the QTM is defined in terms of the
  unitary operator $U_\Delta$, its configurations are
  well-defined only at time $t_n$, for $n\in\NN$.
  The quantum configuration $\ket{\psi_{n+\epsilon}}$ 
  is therefore unknown and depends on how the unitary operator $U_\Delta$ is
  physically realised. In any case,  in general it differs from
  $\ket{\psi_{n}}$. \item $\epsilon =1$. In this case
  the second measurement takes place at time $t_{n+1}$, on the configuration
  $\ket{\psi_{n+1}}=U_\Delta \ket{\psi_n}$. That implies
  $\ket{\psi_{n}}\neq \ket{\psi_{n+1}}$, since  $\ket{\psi_{n}}=\ket{\psi_{n+1}}$
  may hold only in the trivial case the unitary operator
  $U_{\Delta}$ is the identity and the $\ket{\psi_n}=\ket{\psi_0}$, for every $n$.
\end{enumerate}
As a consequence, without additional restrictions on its evolution
operator, Deutsch's approach does not allow to extract the result of
the computation of a QTM.

\subsection{Fixing the measurement protocol}

Ozawa was the first to notice the above problem. In 
\cite{Ozawa-PRL-98}, he gave a revised formulation of the measurement
protocol:
  \begin{quote}\it
  Once the halt qubit is set to the state $\ket{T=1}$, the
  quantum Turing machine no longer changes the halt qubit or the tape
  string.
\end{quote}
But, since such a constraint would be too strong ( as already
remarked, if $\ket{\psi_{n+\Delta}}=\ket{\psi_{n}}$, then $U_\Delta$
must be the identity), Ozawa adds the  footnote 11:
 
\begin{quote}
  \it %
  If the relevant outcome of the computation is designed
  to be written by the program in a restricted part of the tape, this
  condition can be weakened so that the quantum Turing machine may
  change the part of the tape string except that part of the tape.
\end{quote}
In other words, instead of requiring that all tape
stays unchanged after reaching a final configuration, only the
interesting part of the output must become stable after the halt qubit
is set to $1$.

Unfortunately, Linden and Popescu maybe skipped that note, and in
their unpublished paper \cite{LinPop98} claimed that even Ozawa's was
unsatisfactory. So, a few years later, Ozawa replied harshly in
\cite{Ozawa-HQTM-2002}:

\begin{quote} \it
  Recently, Linden and Popescu \cite{LinPop98} claimed that the halt
  scheme given in \cite{Ozawa-PRL-98} is not consistent with unitarity
  of the evolution operator. However, their argument applies only to
  the special case in which the whole tape is required not to change
  after the halt. As suggested in footnote~11 of
  \cite{Ozawa-PRL-98}, the conclusion in \cite{Ozawa-PRL-98} can be
  obtained from the weaker condition for the general case where the
  tape is allowed to change except for the data slot. Linden and
  Popescu \cite{LinPop98} disregarded this case and hence their
  conclusion is not generally true.
\end{quote}

Although Ozawa was right, he did not explain how to stabilise what he
called the data slot. The approach that we shall present in the
following is then a sort of implementation of  Ozawa's protocol,
for which we also prove in details the irrelevance of the number of
executed measurements  and of the measurement time.

\subsection{Our approach}
\label{ssec:our-approach-output}

The measurement of the output computed by our QTMs can be performed by
following a variant of Deutsch's approach. Because of the particular
structure of the transition function of our QTMs, we shall see that we
do not need any additional termination bit, that a measurement can be
performed at any moment of the computation, and that indeed we can
perform several measurements at distinct points of the computation
without altering the result (in terms of the probabilistic
distribution of the observed output).

Given a q-configuration $\ket{\phi}=\ket{\phi_f}+\ket{\phi_{nf}}$,
where $\ket{\phi_f}\in\ell^2(\FConf)$ and
$\ket{\phi_{nf}}\in\ell^2(\GConf\setminus\FConf)$, our
\emph{output measurement} tries to get an output value from
$\ket{\phi}$ by the following procedure:
\begin{enumerate}
\item first of all, we observe the final states of $\ket{\phi}$,
  forcing the q-configuration to collapse either into the final
  q-configuration $\ket{\phi_f}/\norm{\ket{\phi_f}}$, or into the
  q-configuration $\ket{\phi_{nf}}/\norm{\ket{\phi_{nf}}}$, which does not
  contain any final configuration;
\item then, if the q-configuration collapses into
  $\ket{\phi_f}/\norm{\ket{\phi_f}}$, we observe one of these configurations,
  say $\ket{C}$, which gives us the observed output $\val[C]=n$,
  forcing the q-configuration to collapse into the final base q-configuration
  $(e_c/|e_c|)\ket{C}$;
\item otherwise, we leave unchanged the q-configuration
  $\ket{\phi_{nf}}/\norm{\ket{\phi_{nf}}}$ obtained after the first
  observation, and we say that we have observed the special value
  $\bot$.
\end{enumerate}

Summing up, an output measurement of $\ket{\phi}$ may lead to observe
an output value $n\in\NN$ associated to a collapse into a base final
configuration $\ket{C}\in\ket{\phi}$ s.t.\ $\val[\phi]=n$ or to
observe the special value $\bot$ associated to a collapse into
a q-configuration which does not contain any final configuration.

\begin{Definition}[output observation]\label{def:out-obs}
  An \emph{output observation} with collapsed q-configuration
  $\ket{\psi}$ and \emph{observed output} $x\in\NN\cup\{\bot \}$ is
  the result of an output measurement of the q-configuration
  $\ket{\phi}=\sum_{C\in\GConf}e_C\ket{C}$. Therefore, it is a triple
  $\outobs{\phi}{x}{\psi}$ s.t.\
  \begin{enumerate}
  \item either $x=n\in\NN$, and
    \[
      \ket{\psi}=\frac{e_C}{|e_C|}\ket{C}
      \qquad\mbox{ with }\qquad
      C\in\FConf
      \mbox{ and }
      \val[C]=n
      \mbox{ and }
      e_c \neq 0
    \]
  \item or $x=\bot$, and
    \[
      \ket{\psi}=\frac{\ket{\phi_{nf}}}{\norm{\ket{\phi_{nf}}}}
      \qquad\mbox{ where }\qquad
      \ket{\phi_{nf}}=\sum_{C\not\in\FConf}e_C\ket{C}
      \mbox{ and }
      \ket{\phi_{nf}} \neq 0
    \]
  \end{enumerate}
  The \emph{probability of an output observation} is defined by
  \[
    \pr{\outobs{\phi}{x}{\psi}} =
    \begin{cases}
      |e_C|^2 & \qquad \mbox{if } x=n\in\NN\\[+1ex]
      \norm{\ket{\phi_{nf}}}^2 & \qquad \mbox{if } x=\bot
    \end{cases}
  \]
\end{Definition}

\begin{Remark}\label{rem:obs-base-final}
  For \change{$e\in\CC$}, $e\outobs{C}{x}{\phi}$, with $C\in\FConf$ and $\val[C]=n$. Since
  $e\ket{C}$ is a q-configuration, $|e|=1$.  By the definition of
  output observation, $x=n$ and $\ket{\phi}=(e/|e|)U\ket{C}=e\ket{D}$,
  with $\ket{D}=U\ket{C}\in\cb{\FConf}$ and $\val[D]=n$. Moreover,
  $\pr{e\outobs{C}{x}{\phi}}=|e|^2=1$.
\end{Remark}

\begin{Remark}\label{rem:ortonorm-obs}
  For every pair $\outobs{\phi}{x_1}{\psi_1}$ and
  $\outobs{\phi}{x_2}{\psi_2}$ of distinct output observations,
  $\psi_1$ and $\psi_2$ are in the orthonormal subspaces generated by
  the two disjoint sets $\B_1,\B_2\subseteq\GConf$, where
  $\B_i=\{C\in\GConf\mid \ket{C}\in\ket{\psi_i}\}$.
\end{Remark}

\begin{Definition}[observed run]
  Let $M$ be a QTM and $U_M$ be its time evolution operator. For any
  monotone increasing function $\tau:\NN\to\NN$ (that is,
  $\tau(i)<\tau(j)$ for $i<j$):
  \begin{enumerate}
  \item a \emph{$\tau$-observed run} of $M$ on the initial
    q-configuration $\ket{\phi}$ is a sequence
    $\{\ket{\phi_i}\}_{i \in \NN}$ s.t.:
    \begin{enumerate}
    \item $\ket{\phi_0} = \ket{\phi}$;
    \item $U_M\outobs{\phi_{h}}{x_i}{\phi_{h+1}}$, when $h=\tau(i)$ for
      some $i\in\NN$;
    \item $\ket{\phi_{h+1}}=U_M\ket{\phi_{h}}$ otherwise.
    \end{enumerate}
  \item A \emph{finite $\tau$-observed run} of length $k$ is any
    finite prefix of length $k+1$ of some $\tau$-observed run. Notation:
    if $R=\{\ket{\phi_i}\}_{i \in \NN}$, then
    $R[k] = \{\ket{\phi_i}\}_{i \leq k}$.
  \end{enumerate}
\end{Definition}

  In particular, let $id$ be the identity function on
  $\NN$. In the corresponding $id$-observed run, a measurement of the
  tape is performed after each step of the machine.

\begin{Remark}\label{rem:obs-out-run}
  We stress that, given a $\tau$-observed run $R=\{\ket{\phi_i}\}_{i \in \NN}:$
  \begin{enumerate}
  \item either it never obtains a value $n\in\NN$ as the result of an
    output observation, and then it never reaches a final
    configuration;
  \item or it eventually obtains such a value collapsing the
    q-configuration into a base final configuration $e\ket{C}$ s.t.\
    $|e|=1$ and $\val[C]=n$, and from that point onward all the
    configurations of the run are base final configurations
    $e\ket{C_j}=e\,U^j\ket{C}$ s.t.\ $\val[C_j]=n$, and all the
    following observed outputs are equal to $n$ (see
    Remark~\ref{rem:obs-base-final}).
  \end{enumerate}
\end{Remark}
\begin{Definition}
  Let $R=\{\ket{\phi_i}\}_{i \in \NN}$ be a $\tau$-observed run.
  \begin{enumerate}
  \item The sequence $\{x_i\}_{i\in\NN}$ s.t.\
    $\outobs{\phi_{h}}{x_i}{\phi_{h+1}}$, with $h=\tau(i)$, is the
    \emph{output sequence} of the $\tau$-observed run $R$.
  \item The \emph{observed output} of $R$ is the value
    $x\in\NN\cup\{\bot\}$ (notation: $\obsout{R}{x}$) defined by:
    \begin{enumerate}
    \item $x=n\in\NN$, if $x_i=n$ for some $i\in\NN$;
    \item $x=\bot$ otherwise.
    \end{enumerate}
  \item For any $k$, the output sequence of the finite $\tau$-observed
    run $R[\tau(k)+1]$ is the finite sequence $\{x_i\}_{i\leq k}$
    and $x_k$ is its observed output.
  \end{enumerate}
\end{Definition}

\begin{Definition}[probability of a run]
  Let $R=\{\ket{\phi_i}\}_{i \in \NN}$ be a $\tau$-observed run.
  \begin{enumerate}
  \item For $k\in\NN$, the probability of the finite $\tau$-observed
    run $R[k]$ is inductively defined by
    \begin{enumerate}
    \item $\pr{R[0]} = 1$;
    \item $\pr{R[k+1]} = 
      \begin{cases}
        \pr{R[k]}\,\pr{\outobs{\phi_k}{x_i}{\phi_{k+1}}}
        & \
        \parbox{16ex}{%
          when $k = \tau(i)$ for some $i\in\NN$ 
        }
        \\[+1.5em]
        \pr{R[k]} & \ \mbox{otherwise }
      \end{cases}$
    \end{enumerate}
    \medskip
  \item $\pr{R} = \lim_{k\to\infty}\pr{R[k]}$. 
  \end{enumerate}
\end{Definition}

We remark that $\pr{R}$ is well-defined, since
$1\geq \pr{R[i]} \geq \pr{R[j]} > 0$, for every $i\leq j$. Therefore,
$$1\geq \pr{R} = \lim_{k\to\infty}\pr{R[k]}=\inf\{\pr{R[k]}\}_{k\in\NN}\geq 0.$$

\begin{Remark}\label{rem:pr-runs}
  Let $R=\{\ket{\phi_i}\}_{i\in\NN}$ be a $\tau$-observed run s.t.\
  $\obsout{R}{n}$, for some $n\in\NN$. As observed in
  Remark~\ref{rem:obs-out-run}, for some $k$, we have
  $\obsout{R[\tau(k)]}{n}$ and $\obsout{R[\tau(j)]}{\bot}$, for $j<k$;
  moreover, for $i > \tau(k)$, $\ket{\phi_i}=e\ket{C_i}$ with $|e|=1$,
  $C_i\in\FConf$, and $\val[C_i]=n$. As a consequence,
  $\pr{R[\tau(k)+1]}=\pr{R[i]}$, for any $i>\tau(k)$ (since, by
  Remark~\ref{rem:obs-base-final},
  $\pr{\outobs{\phi_{\tau(i)}}{n}{\phi_{\tau(i)+1}}}=1$) and
  $\pr{R}=\lim_{i\to\infty}R[i]=\pr{R[\tau(k)+1]}$.
\end{Remark}

\begin{Definition}[observed computation]\label{def:obs-computation}
  The \emph{$\tau$-observed computation} of a QTM $M$ on the initial
  q-configuration $\ket{\phi}$, is the set
  $\K_{\ket{\phi},\tau}^M$ of the $\tau$-observed runs of $M$
  on $\ket{\phi}$ with the measure
  $\PR:\varP(\K_{\ket{\phi},\tau}^M)\to\CC$
    (where $\varP(\K_{\ket{\phi},\tau}^M)$ is the set of the subsets of $\K_{\ket{\phi},\tau}^M$)
  defined by
  $$\Pr\B = \sum_{R\in\B}\pr{R}$$
  for every $\B\subseteq\K_{\ket{\phi},\tau}^M$.
\end{Definition}

By $\K[k]_{\ket{\phi},\tau}^M$ we shall denote the
set of the finite $\tau$-observed runs of length $k$ of $M$ on
$\ket{\phi}$, with the measure $\PR$ on its subsets (see
Definition~\ref{def:obs-computation}).

It is immediate to observe that the set $\K_{\ket{\phi},\tau}^M$
naturally defines an infinite tree labelled with q-configurations,
where each infinite path starting from the root $\ket{\phi}$
corresponds to a $\tau$-observed run in $\K_{\ket{\phi},\tau}^M$.

\begin{Lemma}
  \label{lem:ortonorm-runs}
  Given $R_1,R_2\in\K_{\ket{\phi},\tau}^M$ \ with
  $R_1=\{\phi_{1,i}\}_{i\in\NN}\neq \{\phi_{2,i}\}_{i\in\NN}=R_2$,
  there is $k\geq 0$ s.t.\
  \begin{enumerate}
  \item\label{item:ortonorm-runs:pre} $\phi_{1,i}=\phi_{2,i}$ for
    $i\leq \tau(k)$, that is, $R_1[\tau(k)]=R_2[\tau(k)]$;
  \item\label{item:ortonorm-runs:post} for $i>\tau(k)$, the
    q-configurations $\phi_{1,i}\neq \phi_{2,i}$ are in two
    orthonormal subspaces generated by two distinct subsets of
    $\GConf$.
  \end{enumerate}
\end{Lemma}
\begin{proof}
  Let $R_1[h]=R_2[h]$ be the longest common prefix of $R_1$ and
  $R_2$. Since they both starts with $\ket{\phi}$, such a prefix is
  not empty; moreover, by the definition of $\tau$-observed run, it is
  readily seen that $h=\tau(k)$, for some $k$.

  Let us now prove item~\ref{item:ortonorm-runs:post}.  If we take
  $\B_{a,j}=\{C\in\GConf\mid \ket{C}\in\ket{\psi_{a,h+1+j}}\}$, for
  $a=1,2$ and $j\in\NN$, we need to prove that
  $\B_{1,j}\cap \B_{2,j}=\emptyset$, for every $j$. By construction,
  $\phi_{1,h+1}\neq \phi_{2,h+1}$ and, for $a=1,2$,
  $\obsout{\phi_h}{x_a}{\phi_{a,h+1}}$, for some $x_a$, with
  $\phi_h=\phi_{a,h}$. Moreover, $\B_{1,0}\cap \B_{2,0}=\emptyset$
  (see Remark~\ref{rem:ortonorm-obs}), at least one of the two
  observed values $x_1,x_2$ is not $\bot$, and one of the two
  q-configurations $\ket{\phi_{1,h+1}},\ket{\phi_{2,h+1}}$ is a final
  base q-configuration. W.l.o.g., let us assume that $x_1\in\NN$ and
  let $\ket{\phi_{1,h+1}}=e\ket{C,n}$, for some final plain
  configuration $C$ and some $n\in\NN$.  Then,
  $\ket{\phi_{1,h+1+j}} = e\,U^j\ket{C,n}=e\ket{C,n+j}$ (see
  Remark~\ref{rem:obs-out-run}) and $\B_{1,j} =\{\ket{C,n+j}\}$, for
  $j\in\NN$. Thus, to prove the assertion, it suffices to show that
  $\ket{C,n+j}\not\in\B_{2,j}$, for $j\in\NN$. Let us distinguish two
  cases:
  \begin{enumerate}
  \item $x_2\in\NN$ and $\ket{\psi_{2,h+1}}=e'\ket{D,m}$, where $D$ is
    a final plain configuration and $m\in\NN$. As already seen for
    $\ket{\psi_{1,h+1+j}}$, for $j\in\NN$, we have
    $\ket{\psi_{2,h+1+j}}=e'\,U^j\ket{D,m}=e'\ket{D,m+j}$ and
    $\B_{2,j} =\{\ket{D,m+j}\}$. Therefore,
    $\ket{C,n+j}\not\in\B_{2,j}$, since $\ket{C,n+j}\neq\ket{D,m+j}$,
    for $\ket{C,n}\neq\ket{D,m}$ by construction.
  \item $\ket{\psi_{2,h+1}}\in\ell^2(\GConf\setminus \FConf)$.  Let
    $\ket{D,m}\in\B_{2,j}\cap\FConf$, where $D$ is a plain
    configuration and $m\in\NN$.  By induction on $j$, it is readily
    seen that $m < j \leq n+j$. Thus, $\ket{D,m}\neq \ket{C,n+j}$.
  \end{enumerate}
\end{proof}


\begin{Lemma}\label{lem:obs-reconstr-qconf}
  Let $K _{\ket{\phi}}^M=\{\phi_i\}_{i\in\NN}$ be the computation
  of the QTM $M$ on the initial q-configuration $\ket{\phi}$ and
  $\K_{\ket{\phi},\tau}^M$ be the $\tau$-observed computation on
  the same initial configuration.  For every $k\in\NN$, we have
  that
  \[
    \ket{\phi_k} = \sum_{R\in\K[k]_{\ket{\phi},\tau}^M} \sqrt{\pr{R}}\,\ket{\psi_{R}}
  \]
  where $\ket{\psi_R}$ is the last q-configuration of the finite run
  $R$ of length $k$.
\end{Lemma}
\begin{proof}
  By definition, $\phi=\phi_0$ and $R=\{\phi\}$ with $\pr{R}=1$, and
  $\psi_R=\phi$ is the only run of length $0$ in
  $\K_{\ket{\phi},\tau}^M$. Therefore, the assertion
  trivially holds for $k=0$. 

  Let us then prove the assertion by induction on $k$. By definition
  and the induction hypothesis
  $$\ket{\phi_{k+1}} = U_M\ket{\phi_k} = 
  \sum_{R\in\K[k]_{\ket{\phi},\tau}^M} \sqrt{\pr{R}}\,U_M\ket{\psi_R}$$
  We have two possibilities:
  \begin{enumerate}
  \item $k\neq\tau(i)$, for any $i$. In this case, there is a bijection
    between the runs of length $k$ and those of length $k+1$, since
    each run $R'\in\K[k+1]_{\ket{\phi},\tau}^M$ is obtained
    from a run $R\in\K[k]_{\ket{\phi},\tau}^M$ with last
    q-configuration $\ket{\psi_R}$, by appending to $R$ the
    q-configuration $\ket{\psi_{R'}}=U_M\ket{\psi_R}$. Moreover, since
    by definition, $\pr{R'}=\pr{R}$, we can conclude  that
    \[
      \ket{\phi_{k+1}} = 
      \sum_{R\in\K[k]_{\ket{\phi},\tau}^M} \sqrt{\pr{R}}\,U_M\ket{\psi_R} =
      \sum_{R'\in\K[k+1]_{\ket{\phi},\tau}^M} \sqrt{\pr{R'}}\ket{\psi_{R'}}
    \]
    \item $k=\tau(i)$, for some $i$. In this case, every
    $R\in\K[k]_{\ket{\phi},\tau}^M$ with last q-configuration
    $\ket{\psi_R}$ generates a run $R'$ of length $k+1$ for every
    output observation $\outobs{\psi_R}{x}{\psi_{R'}}$, where $R'$ is
    obtained by appending $\ket{\psi_{R'}}$ to $R$. Therefore, let
    $R=\{\ket{\psi_i}\}_{i\leq k} \in
    \K[k]_{\ket{\phi},\tau}^M$ and
    $$\B_R = \{ \{\psi_i\}_{i\leq k+1} \mid
    \outobs{\psi_k}{x}{\psi_{k+1}}\}$$
    By applying Definition~\ref{def:out-obs}, we easily check that
    \[
      U_M\ket{\psi_R} =
      \sum_{R'\in\B_R}\sqrt{\pr{\outobs{\psi_R}{x}{\psi_{R'}}}}\,\ket{\psi_{R'}}
    \]
    Thus, by substitution, and
    $\pr{R}\,\pr{\outobs{\psi_R}{x}{\psi_{R'}}}=\pr{R'}$
    \begin{align*}
      \ket{\phi_{k+1}} 
      & = 
        \sum_{R\in\K[k]_{\ket{\phi},\tau}^M} \sqrt{\pr{R}}\,U_M\ket{\psi_R} 
      \\
      & =
        \sum_{R\in\K[k]_{\ket{\phi},\tau}^M}
        \sum_{R'\in\B_R}
        \sqrt{\pr{R}}\,\sqrt{\pr{\outobs{\psi_R}{x}{\psi_{R'}}}}\,\ket{\psi_{R'}}
      \\
      & = 
        \sum_{R'\in\,\bigcup_{R\in\K[k]_{\ket{\phi},\tau}^M}\B_R} \sqrt{\pr{R'}}\,\ket{\psi_{R'}}
      \\
      & = 
        \sum_{R'\in\K[k+1]_{\ket{\phi},\tau}^M} \sqrt{\pr{R'}}\,\ket{\psi_{R'}}
    \end{align*}
    since
    $\bigcup_{R\in\K[k]_{\ket{\phi},\tau}^M}\B_R=\K[k+1]_{\ket{\phi},\tau}^M$.
  \end{enumerate}
\end{proof}

We are finally in the position to prove that our observation protocol
is compatible with the probability distributions that we defined as
computed output of a QTM computation.

\begin{Theorem}\label{thm:probabilities}
  Let $K _{\ket{\phi}}^M=\{\phi_i\}_{i\in\NN}$ be the computation
  of the QTM $M$ on the initial q-configuration $\ket{\phi}$ and
  $\K_{\ket{\phi},\tau}^M$ be the $\tau$-observed computation on
  the same initial configuration. For every $n\in\NN$:
  \begin{enumerate}
  \item
    $\pd{\ket{\phi_k}}(n) = \pr{R\in
      \K[k]_{\ket{\phi},\tau}^M \mid \obsout{R}{n}}$,
    for $k=\tau(i)+1$ and $i\in\NN$;
  \item $\pd{K _{\ket{\phi}}^M}(n) = \pr{R\in \K_{\ket{\phi},\tau}^M \mid \obsout{R}{n}}$.
  \end{enumerate}
\end{Theorem}
\begin{proof}
  Let us start with the first item.  By
  Lemma~\ref{lem:obs-reconstr-qconf}, we know that
  $$\ket{\phi_k} = \sum_{R\in\K[k]_{\ket{\phi},\tau}^M}
  \sqrt{\pr{R}}\,\ket{\psi_{R}},$$
  where $\ket{\psi_R}$ is the last
  q-configuration of $R$. Since $k=\tau(i)+1$, for some $i$, we also
  know that either $\psi_R\in\GConf\setminus\FConf$ or
  $\ket{\psi_R}=u_R\ket{C_R}$ with $C_R\in\FConf$ and
  $|u_R|=1$. Therefore,
  $$\pd{\ket{\phi_k}}(n)=
  \norm{\sum_{R\in\B[k,n]} \sqrt{\pr{R}}\,u_R\ket{C_{R}}}^2$$
  where
  \begin{align*}
    \B[k,n] 
    & = 
      \{R\in \K[k]_{\ket{\phi},\tau}^M \mid \obsout{R}{n}\}
    \\
    & = 
      \{R\in\K[k]_{\ket{\phi},\tau}^M \mid
      \ket{\psi_R}=u_R\ket{C_R} \mbox{ with } \val[C_R]=n\}
  \end{align*}
  
  By Lemma~\ref{lem:ortonorm-runs}, we know that for every
  $R_1,R_2\in\B[k,n]$, we have
  $\ket{C_{R_1}}\neq\ket{C_{R_2}}$. Therefore
  $$\pd{\ket{\phi_k}}(n)=
  \sum_{R\in\B[k,n]} \pr{R}|\,u_R|^2 =
  \sum_{R\in\B[k,n]} \pr{R} =\ \PR \B[k,n]$$
  since $|u_R|=1$. Which concludes the proof of the first item of the
  assertion.

  In order to prove the second item, let
  $\B[\omega,n] = \{R\in \K_{\ket{\phi},\tau}^M \mid
  \obsout{R}{n}\}$. We have that
  \begin{multline*}
    \pr{R\in \K_{\ket{\phi},\tau}^M \mid \obsout{R}{n}}
    = \sum_{R\in \B[\omega,n]}\pr{R}
    = \lim_{i\to\infty} \sum_{\dind{R\in\B[\omega,n]}{\obsout{R[\tau(i)+1]}{n}}} \pr{R}
    \\ 
    {}^{(*)}= \lim_{i\to\infty} 
        \sum_{\dind{R\in\B[\omega,n]}{\obsout{R[\tau(i)+1]}{n}}} \pr{R[\tau(i)+1]}
    \\
    {}^{(**)}= \lim_{i\to\infty}\sum_{R'\in \B[\tau(i)+1,n]}\pr{R'}
    = \lim_{i\to\infty} \PR\, \B[\tau(i)+1,n]
  \end{multline*}
  since: $(*)$ $\pr{R}=\pr{R[\tau(i)+1]}$, when $\obsout{R[\tau(i)+1]}{n}$ (see
  Remark~\ref{rem:pr-runs}); $(**)$ there is a bijection between 
  $\B[\tau(i)+1,n]$ and
  $\{R\in \B[\omega,n] \mid \obsout{R[\tau(i)+1]}{n}\}$ (see
  Remark~\ref{rem:obs-out-run}) mapping every
  $R'\in\B[\tau(i)+1,n]$ with last q-configuration $u\ket{C}$
  into
  $R = \{\psi_j\}_{j\in\NN}\in \{R\in \B[\omega,n]
  \mid \obsout{R[\tau(i)+1]}{n}\}$
  s.t\ $R'=R[\tau(i)+1]$ and
  $\ket{\psi_j}=u\,U_M^{j-\tau(i)-1}\ket{C}$ for $j> \tau(i)$.
  
  Therefore, by the (already proved) first item of the assertion
  $$\pr{R\in \K_{\ket{\phi},\tau}^M \mid \obsout{R}{n}}
  = \lim_{i\to\infty}\PR\, \B[\tau(i)+1,n]
  = \lim_{i\to\infty}\pd{\ket{\phi_{\tau(i)+1}}}(n)
  = \pd{K _{\ket{\phi}}^M}(n)$$
\end{proof}

The above theorem shows that our protocol is not only
  realistic (unlike those of \BeV and Deutsch), but also compatible with the
  theoretical probabilities defined for configurations in
  superposition. In particular, the theorem shows how it is possible
  to make measurements in a totally free way, without destroying the expressive power
  of quantum parallelism, i.e., without destroying the ability of a QTM
  to use quantum interference.

\section{A comparison with Bernstein and Vazirani's QTMs: part 2}
\label{sec:comp-BandV-2}

Theorem~\ref{theor:fromBeVtoOur}, has shown that  our QTMs
generalise \BeV-QTMs, since the computation of any \BeV-QTM can be
simulated by a corresponding QTM with the same transition function (up
to transitions entering/leaving the initial/final state). The general
framework, however, is substantially modified, and the ``same''
machine behaves in different ways in the two approaches. In this
section, we give two simple examples of this.

\begin{figure}[!htb]
  \begin{center}
    \scalebox{0.70}{%
      \includegraphics{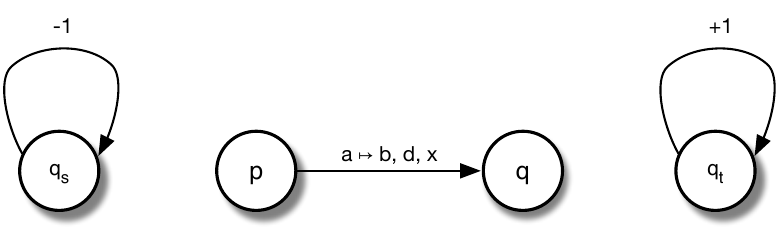} 
    }
  \end{center}
  \caption{Transitions of a QTM}\label{fig:transitions}
\end{figure}

\subsection{QTM transition graphs}
\label{sec:qtm-trans-graphs}

Let us represent a QTM $M$ by means of a \emph{transition graph}
(a directed graph) whose nodes are the states of $M$, and whose
arrows give its transition function (see
Figure~\ref{fig:transitions}). Namely, if
$\delta_0(p,a)(q,b,d)=x\neq 0$, the graph of $M$ contains an arrow
from the node of $p$ to the node of $q$, labelled by the tuple
$(a \mapsto b, d, x)$. Every non-target node has at least one outgoing
edge labelled by such a tuple, for any symbol $a$ of the tape
alphabet.

In addition to the arrows of $\delta_0$, to represent the looping
transitions of target nodes, the graph contains a self-loop labelled
by $+1$ on any target node $q_t$: to denote the fact that the only
transition of any target configuration is the one that increases the
counter by $1$, without modifying the rest of the
configuration. Dually, every source node $q_s$ has a self-loop
labelled by $-1$. The self-loop of $q_s$ is not its only outgoing
arrow, since the corresponding counter decreasing transition applies
to source configurations with a counter greater than $0$ only; indeed,
when the counter of a source configuration reaches the $0$, the
$\delta_0$ transition function applies. On the other hand, the
self-loop is the only incoming arrow of any source state $q_s$, since
no transition from another state can enter into it. The situation is
dual for target states, for which the self-loop is the only outgoing
arrow.

See Figure~\ref{fig:ide} for an example of source and target nodes
of a transition graph: the state $s$ and $q_0$ (which is also initial)
are source states; the states $p$ and $q_f$ (which is also final) are
target states.

\subsection{A classical reversible TM with quantum behaviour}\label{Section:reversibleTM}

\begin{figure}[!htb]
  \begin{center}
    \scalebox{0.70}{%
      \includegraphics{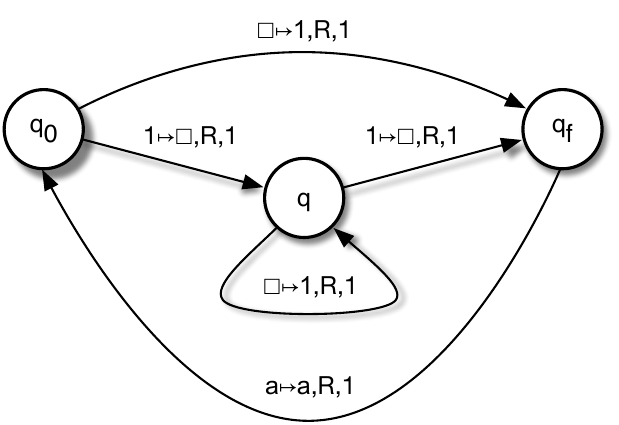} 
    }
  \end{center}
  \caption{A reversible TM as a QTM à la Bernstein and Vazirani}
  \label{fig:RTMBeV}
\end{figure}

In Figure~\ref{fig:RTMBeV}, we give the transition graph of a \BeV-QTM
corresponding to a reversible TM: any reversible TM $M$ can be
transformed into a \BeV-QTM by assuming that the weight of any
transition of $M$ is $1$, and by adding a back-transition from the
final state to the initial state.

In the example in Figure~\ref{fig:RTMBeV}, $q_0$ is the initial state
and $q_f$ is the final one, and they are connected by an arrow
$(a\mapsto a, R,1)$ from $q_f$ to $q_0$, where $a\in\{\Box,1\}$.  When 
started on an initial tape containing \change{%
$\nstring{n+1}$ (that is, $n+2$ symbols $1$, see Section~\ref{sec:qtm}), the machine $M$
erases two $1$s from the tape. When the initial tape contains instead 
$\nstring{0}$, $M$ loops indefinitely on the middle
state $q$, after erasing the unique symbol $1$ on the tape.  Summing
up (remembering that for the computed output we just count the number of $1$s
on the tape, see Definition~\ref{def:prob-comf}), 
$M$ computes the predecessor $n-1$ (with probability $1$), when feeded with a representation of
$n>0$, while it diverges (with probability $1$) for $n=0$.
}
However, if we
feed $M$ with a non classical input as
$\ket{\psi} = \frac{1}{\sqrt{2}}\ket{\nstring{0}}+
\frac{1}{\sqrt{2}}\ket{\nstring{2}}$, then $M$ fails to give an answer
according to \BeV's framework, since it reaches a q-configuration in
which final and non-final base configurations superpose.

\begin{figure}[!htb]
  \begin{center}
    \scalebox{0.70}{%
      \includegraphics{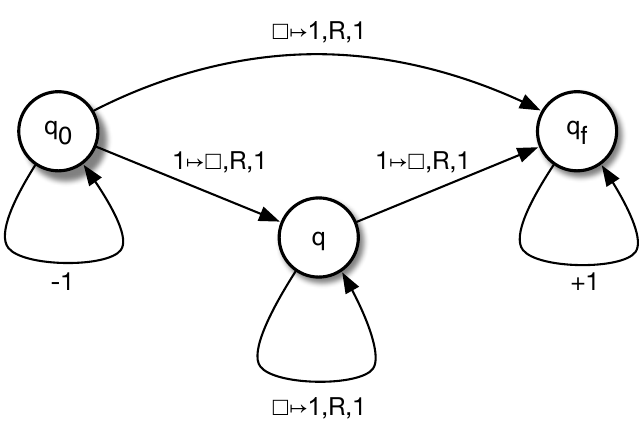} 
    }
  \end{center}
  \caption{A reversible TM as a QTM}\label{fig:RTM}
\end{figure}

To transform $M$ into a QTM according to our formalism (see
Theorem~\ref{theor:fromBeVtoOur}), it suffices to replace the arrow
from $q_f$ to $q_0$ by two self-loops: one (labelled $-1$) on the
initial state $q_0$, and one (labelled $+1$) on the final state
$q_f$. We obtain then the QTM in Figure~\ref{fig:RTM}.  By the
definition of computed output, we can see that
$M_{\frac{1}{\sqrt{2}}\ket{\nstring{0}}+\frac{1}{\sqrt{2}}\ket{\nstring{2}}}\to
\{1\mapsto 1/2; n \mapsto 0, \mbox{if } n\neq 1\}$; namely,
with probability $1/2$ the QTM halts with computed output $1$;
while with probability $1/2$ it diverges.

\subsection{A PD obtained as a limit}
\label{ssec:ex:identity}

\begin{figure}[!htb]
  \begin{center}
    \scalebox{0.80}{%
      \includegraphics{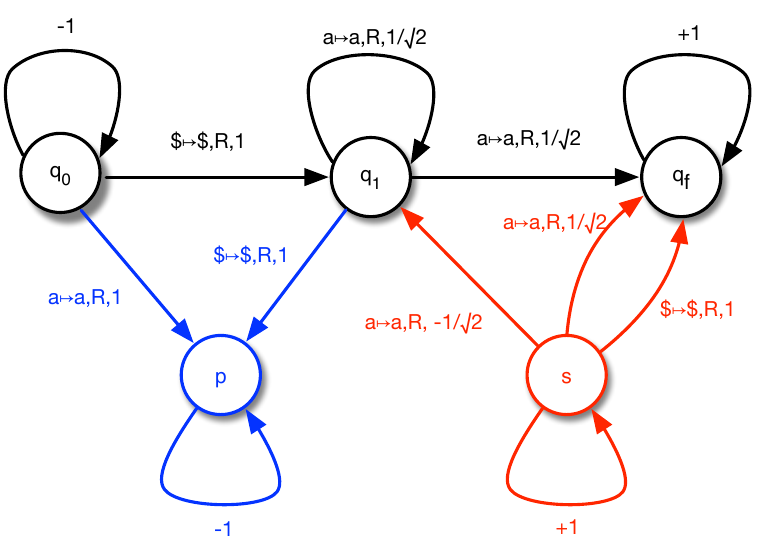} 
    }
  \end{center}
  \caption{A QTM computing the successor function (the identity
    w.r.t.\ the tape) as a limit}
  \label{fig:ide}
\end{figure}

The example in Figure~\ref{fig:ide} shows a QTM which produces a PD
only as an infinite limit.  The tape alphabet of the machine $M$ is
$\Sigma=\{\$,1,\Box\}$, the set of its source states is
$Q_s=\{q_0,s\}$, the set of its target states is $Q_t=\{q_f,p\}$, the
state $q_0$ is initial, the state $q_f$ is final. In the figure, the
symbol $a$ stands for any symbol of the alphabet but $\$$, that is,
$a\in\{1,\Box\}$.

The machine $M$ applies properly on initial configurations $\langle
\lambda, q_0, \$\nstring{n}, 0, 0
\rangle$, whose corresponding base vector will be denoted by
$\ket{\$\nstring{n}}$ (to simplify the definition of the machine, we
use a slight variant of the enconding given in
subsection~\ref{ssec:quant-part-comp-fun} in which the sequence
$\nstring{n}$ coding the number $n$ is preceded by a $\$$, which is
indeed the only $\$$ on the tape).  On such inputs, after moving from
the initial state $q_0$ to $q_1$,
$M$ either ends up in the final state $q_f$ with probability
$1/2$, or it loops on $q_1$ with probability
$1/2$. We remark the source state $s$ and the target state
$p$: such states do not play any role when the machine computes on
$\$\nstring{n}$, since none of them can be reached. Nevertheless, they
must be added to deal with error situations, and to satisfy the local
unitary conditions, in order to get a proper unitary time evolution
for $M$ \change{(see Remark~\ref{rem:source-target})}.

A simple calculation shows that $M_{\ket{\$\nstring{n}}}\to\pd{}$,
with $\pd{} = \{n+1\mapsto 1; m\mapsto 0, \mbox{if } m \neq n+1\}$;
namely, on the input $\$\nstring{n}$, $M$ computes with probability
$1$ the successor $n+1$. (The machine is indeed the identity w.r.t.\
the tape, since no transition of the machine changes any symbol on the
tape; however, since $\nstring{n}$ is a sequence of $n+1$ symbols $1$,
the computation leaves all these symbols $1$ on the tape, leading then
to a computed output of $n+1$.)  We stress that the PD $\pd{}$ is
obtained as a limit, since for every $j\in\NN$,
$\pd{U^j\ket{\$\nstring{n}}}$ is a PPD s.t.\
$\pd{U^j\ket{\$\nstring{n}}}(n+1)<1$ and
$\pd{U^j\ket{\$\nstring{n}}}(m)=0$, for $m\neq n+1$. Of course, this
does not mean that we have to wait an infinite time to readback the
result!  A correct way to interpret this fact is that for each
$n\in\NN$, and each $\epsilon\in (0,1/2]$, there exists $k\in\NN$ s.t,
for every $j>k$,
$1-\epsilon < \pd{U^j\ket{\$\nstring{n}}}(n+1) \leq 1$.

   
\section{Related works}\label{sect:related}

After having analysed at the length the notions of \BeV and Deutsch,
we discuss in this section some other
papers related to our work.

\subsection{On quantum extensions of Turing Machines, and
  of related complete formalisms}
\label{ssec:quant-extens-TM}

\begin{itemize}
\item Strictly following the \BeV approach, Nishimura and
  Ozawa~\cite{NiOz02,NisOza2009} study the relationship between QTMs
  and quantum circuits (extending previous results
  by Yao~\cite{Yao93}).  They show that there exists a
  perfect computational correspondence between QTMs and uniform,
  finitely generated families of quantum circuits. Such a
  correspondence preserves the quantum complexity classes EQP, BQP and
  ZQP.

\item Perdrix and Jorrand~\cite{Perdr2006} proposes a new way to deal with quantum
  extensions of Turing Machines.  The basic idea is reminiscent of the
  quantum-data/classical-control paradigm coined by
  Selinger~\cite{Sel04c,SelVal06}. In fact, in Perdrix QTM's, the only
  quantum component is the tape whereas the control is completely
  classical.

\item Dal~Lago, Masini, and Zorzi~\cite{DLMZmscs08,DMZ10,DLMZentcs11} extend the
  quantum-data/classical-control paradigm to a type free quantum
  $\lambda$-calculus that is proven to be in perfect correspondence
  with the QTMs of \BeV. Following the ideas of the so called Implicit
  Computational Complexity, the authors propose an alternative way to
  deal with the quantum classes EQP, BQP, and ZQP.
\end{itemize}

\subsection{On the readout problem}
\label{ssec:readout-problem}

The following papers address the problem of \emph{how to readout
  the result of a quantum computation}. Since this is a key question
in the definition of any quantum computing formalism, they
deserve some deeper attention. 

 {We recall, however, that our main
interest is in QTMs as devices computing distributions of probability, and not
functions over natural numbers.}

\begin{itemize}
\item Myers~\cite{Mye97} tries to show that it is not possible to
  define a truly quantum general computer.  The article highlights how the \BeV approach fails on
  truly quantum data.  In fact, in such a case it is impossible to
  guarantee the synchronous termination of all the computations in
  superposition. Consequently, the use of a termination bit spoils the
  quantum superposition of the computation.  This defect was well
  known, and it is for this reason that \BeV did not define a
  general notion of quantum computability, but rather a notion
  sufficient to solve---in a quantum way--- only classical decision 
  problems.  Myers's criticism \emph{does not apply to our approach}.  Our
  QTMs are fully quantum, and they have an observational protocol of
  the result that does not depend on the synchronous termination of
  the computations in superposition.

\item In an unpublished note, Kieu and Danos~\cite{KiDa98} claim that:
  ``\textit{For halting, it is desirable of the dynamics to be able to
    store the output, which is finite in terms of qubit resources,
    invariantly (that is, unchanged under the unitary evolution) after
    some finite time when the desirable output has been computed}.''
  Unfortunately, it is not possible to enter into a truly invariant
  final quantum configuration---only a machine starting in a final
  configuration and computing the identity can accomplish this
  constraint.  We overcome the problem by introducing a feasible 
  (i.e., correct from a quantum point of view) notion of invariant, 
  w.r.t.\ the readout, of final
  configurations. In this way, even if the final configuration
  changes, the output we read from that configuration does not change.

\item In another unpublished note, Linden and Popescu~\cite{LinPop98}
  address the problem of how to readout the result of a general
  quantum computer.  The authors write: ``\textit{We explicitly
    demonstrate the difficulties that arise in a quantum computer when
    different branches of the computation halt at different, unknown,
    times}'', implicitly referring to the problems in extending the
  approach of \BeV to general quantum inputs (see again~\cite{Mye97},
  discussed above).  In the first part of the work, the authors show
  that the problem cannot be solved by means of the so-called
  ``ancilla''.  The ancilla is an additional information added to the
  main information encoded by a configuration of the quantum
  machine. The idea is that, once a final state is reached, the
  machine keeps modifying the ancilla only. The authors show that the
  ancilla approach destroys the quantum capabilities of quantum
  machines, since only classical computations can survive to this
  treatment of ancilla.  Even if reminiscent of the ancilla, our
  approach is technically different---the problems addressed by Linden
  and Popescu do not apply---since we carefully tailor the space of
  the possible configurations of the machines, allowing the ancilla to
  play a role only during its final and initial evolution.

  In the second part of the work, the authors launch a strong attack
  against the use of termination bit, the solution originally proposed
  by Deutsch and successively refined by Ozawa \cite{Ozawa-PRL-98}.
  The authors try to argue that the approach proposed by Deutsch/Ozawa
  cannot work. In fact, they show that even if it is true that once
  the termination bit is set to 1 it remains firmly with such a value
  forever, any terminal configuration cannot be frozen, and keeps
  evolving according to the Hamiltonian of the system. 
  Once again, our proposal does not have the defect depicted in the paper,
  because, far away to force a final configuration to remain stable,
  only the readout of a final configuration is stable in our approach.

\item Hines \cite{Hin10} shows how to ensure simultaneous coherent
  halting, provided that termination is guaranteed for a restricted
  class of quantum algorithms.  This kind of approach based on
  coherent halting is intentionally not followed in our paper. Indeed,
  as previously remarked, we are interested in treating systems which
  includes, as a particular case, all classical computable
  functions---we cannot restrict to terminating
  computations. 

\item Miyadera and Ohya~\cite{MiOh05} discuss the notion of
  \emph{probabilistic halting}. In particular, 
    ``\textit{[\ldots] the notion of halting is still probabilistic. That is,
    a QTM with an input sometimes halts and sometimes does not
    halt. If one can not get rid of the possibility of such a
    probabilistic halting, one can not tell anything certain for one
    experiment since one can not say whether an event of halting or
    non-halting occurred with probability one or just by accident, say
    with probability $10^{-40}$.}'' Therefore, they wonder about the
  existence of any algorithm to decide whether or not a QTM
  probabilistic halts. With no surprise, they conclude that such an
  algorithm cannot exist. In fact, since the non-probabilistic halting
  of a QTM corresponds to the simultaneous halting of all the
  superposing branchings of its computation, an algorithm deciding the
  probabilistic halting would decide the simultaneous termination of
  two classical reversible machines (by combining them into a
  unique QTM), which is clearly an undecidable problem. In a sense,
  the question of probabilistic or non-probabilistic halting is
  irrelevant for our approach. In our QTMs, the result of a
  computation is defined as a limit, and any computation converges to
  some result. Accordingly, the only possible readouts that we can get
  are approximations of such a result. At the same time, since we show
  that a repeated-measures protocol can retrieve the distribution
  associated to the output of a computation, we can accept to say that
  our approach is probabilistic.
\end{itemize}

\section{Conclusions and further work}
\label{sec:concl-furth-work}

We find surprising that in the thirty years since~\cite{Deu85} a
theory of quantum computable functions did not develop, and that the
main interest remained in QTMs as computing devices for classical
problems/functions. This in sharp contrast with the original
(Feynman's and Deutsch's) aim to have a better computing simulation of
the physical world. 

As always in these foundational studies, we had to go back to the
basics, and look for a notion of QTM general enough to encompass
previous approaches.
\change{We started with pre Quantum TMs (Definition~\ref{def:pQTM}), and 
defined then QTMs as those pre QTMs whose time evolution
operator is unitary (Definition~\ref{def:qtm}). Following
analogous results in the literature, we showed that unitarity 
of the evolution operator for a QTM $M$ is equivalent to a set of constraints formulated
on the transition function of $M$ (Theorem~\ref{theor:local}). 
Our QTMs are able to simulate \BeV-QTMs
(Theorem~\ref{theor:fromBeVtoOur}), but, contrary to \BeV-QTMs, they may be used
to give meaning to some ``infinite'' computations. Indeed, in Section~\ref{sec:quant-comp-fun}
we showed that each QTM $M$ naturally defines a computable function from the
sphere of radius $1$ in $\ell^2(\GConf_M)$ (i.e., the superpositions of the configurations of $M$) to the set of (partial)
probability distributions on the set of natural numbers. 
Such a definition is one
of the main reasons for our definition of QTM, for it involves computations where
some  path ``terminates'', while others do not.  
We remark that the definition implies monotonicity
of quantum computations (Theorem~\ref{theor:monot}), which is a consequence of
the particular way final states are treated in order to defuse quantum
interference, once such states are entered. 

Particularly tricky is the way in which one reads a result from 
a quantum computation, which is the subject of Section~\ref{sec:observables}, 
and in particular~\ref{ssec:our-approach-output}, which shows that 
our protocol is compatible with the
theoretical probabilities defined for configurations in
superposition (Theorem~\ref{thm:probabilities}).
}

While several details of
the proposed approach may well change during further study, we are
confident  the general picture 
may allow a fresh look to the problem of a quantum universal machine, 
\change{%
to the termination of QTMs (for instance characterising termination into the arithmetical hierarchy),
}
and to
the various degrees
of partiality of quantum computable functions.

\bibliographystyle{abbrv}
\bibliography{biblio} 

\newpage

\appendix


\section{Hilbert spaces with denumerable basis}
\label{sec:HS}

\begin{Definition}[Hilbert space of configurations]
  Given a denumerable set $\B$, with $\ell^2(\B)$ we shall denote the
  infinite dimensional Hilbert space defined as follow.

  The set of vectors in $\ell^2(\B)$ is the set
  $$
  \left\{\phi\;|\;\phi:\B\rightarrow\CC, \sum_{C\in
      \B}|\phi(C)|^2 < \infty\right\}
  $$ 
  and equipped with:
  \begin{enumerate}
  \item An {inner sum} $+ : \ell^2(\B) \times\ell^2(\B)\to\ell^2(\B)$
    \\
    defined by $(\phi+\psi)(C)= \phi(C)+\psi(C)$;
  \item A {multiplication by a scalar}\quad
    $\cdot:\CC\times\ell^2(\B)\to \ell^2(\B)$
    \\
    defined by $(a\cdot \phi)(C)= a\cdot(\phi(C))$;
  \item An {inner product}\footnote{%
      The condition $\sum_{C\in\B}|\phi(C)|^2<\infty$ implies that
      $\sum_{C\in\B}\phi(C)^*\psi(C)$ converges for every pair of
      vectors.}
    $\inprod{\cdot}{\cdot}: \ell^2(\B)\times\ell^2(\B)\to\CC$\\
    defined by $\inprod{\phi}{\psi}=\sum_{C\in \B}\phi(C)^*\psi(C)$;
  \item The Euclidian norm is defined as
    $\norm{\phi}=\sqrt{\inprod{\phi}{\phi}}$.
  \end{enumerate}
\end{Definition}

The Hilbert space $\ell^2=\ell^2(\NN)$ is the standard Hilbert space
of denumerable dimension---all the Hilbert spaces with denumerable
dimension are isomorphic to it. $\ell^2_1$ is the set of the vectors
of $\ell^2$ with unit norm.

\begin{Definition}[computational basis]
  The set of functions 
  $$\cb{\B}=\{\ket{C} : C\in
  \B,\ \ket{C}: \B\to \CC \}$$
  such that for each $C$
   $$
   \ket{C}(D)= \left\{
     \begin{array}{ll}
     1\;\;&\mbox{if}\;C=D\\
     0\;\;&\mbox{if}\;C\neq D 
     \end{array}
   \right. 
   $$
  is called  \emph{computational basis} of $\ell^2(\B)$.
\end{Definition}

We can prove that~\cite{RomanBook}:
\begin{Theorem}
  The set $\cb{\B}$ is a Hilbert basis of $ \ell^2(\B)$.
\end{Theorem}

Let us note that the inner product space $\SPAN{\cb{\B}}$ defined by:
$$
\SPAN{\cb{\B}} =\left\{\sum_{i=1}^n c_i S_i \ |\ c_i\in\CC, S_i\in \cb{\B}, n\in\NN \right\}.
$$
is a proper inner product subspace of $\ell^2(\B)$, but it is not an
Hilbert Space (this means that $\cb{\B}$ is not an Hamel basis of
$\ell^2(\B)$).

The completion of $\SPAN{\cb{\B}}$ is a space isomorphic to  $ \ell^2(\B)$.

By means of a standard result in functional analysis we have:
\begin{Theorem}
  \mbox{}
  \begin{enumerate}
  \item $\SPAN{\cb{\B}}$ is a dense subspace of
    $\ell^2(\B)$;
  \item 
    $\ell^2(\B)$ is the (unique! up to isomorphism)
    \emph{completion} of $\SPAN{\cb{\B}}$.
  \end{enumerate}
\end{Theorem}

\begin{Definition}
  Let $\V$ be a complex inner product space, a linear
  application $U:\V\to\V$ is called an
  \emph{isometry} if $\inprod{Ux}{Uy}=\inprod{x}{y}$, for each
  $x,y\in \V$; moreover if $U$ is also surjective, then it is
  called \emph{unitary}.
\end{Definition}

Since an isometry is injective, a unitary operator is invertible, and
moreover, its inverse is also unitary.

\begin{Definition}
  Let $\V$ be a complex inner product vector space, a
  linear application $L:\V\to\V$ is called
  \emph{bounded} if $\exists c>0\;\forall x\; \|Lx\|\leq c \|x\|$.
\end{Definition}

\begin{Theorem}
  Let $\V$ be a complex inner product vectorial space, for
  each bounded application $U:\V\to\V$ there is one
  and only one bounded application $U^*:\V\to\V$
  s.t. $\inprod{x}{Uy}=\inprod{U^*x}{y}$. We say that $U^*$ is the
  \emph{adjoint} of $U$.
\end{Theorem}

It is easy to show that if $U$ is a bounded application, then $U$ is
unitary iff $U$ is invertible and $U^*=U^{-1}$.

\begin{Theorem}\label{standard-ext-U}
  Each bounded unitary operator $U$ in
  $\SPAN{\cb{\B}}$ has an unique continuous extension in
  $\ell^2(\B)$~\cite{BerVa97}.
\end{Theorem}

\subsection{Dirac notation}
\label{ssec:dirac-notation}

We conclude this brief digest on Hilbert spaces, by a synopsis of the
so-called Dirac notation, extensively used in the paper.

\bigskip

\begin{center}
  \begin{tabular}{|c|c|}
    \hline
    \textsf{mathematical notion} & \textsf{Dirac notation}\\
    \hline
    inner product $\inprod{\phi}{\psi}$ & $\kinprod{\phi}{\psi}$\\
    \hline
    vector $\phi$ & $\ket{\phi}$
    \\
    \hline
    dual of vector $\phi$ & $\tek{\phi}$\\
    i.e., the linear application $d_\phi$ &
    \\
    defined as $d_\phi(\psi)=\inprod{\phi}{\psi}$ & 
                                                    note that $\kinprod{\phi}{\psi}=\tek{\phi}(\ket{\psi})$
    \\
    \hline
  \end{tabular}
\end{center}

\bigskip

Let $L$ be a linear application, with $\tek{\phi}L\ket{\psi}$ we denote $\kinprod{\phi}{L\psi}$.


\section{Implementation of the counter}
\label{sec:impl-counter}

A TM purist might argue that the counter adds to the machine a device
with a denumerable set of symbols, or with a denumerable set of
states, which is not in the spirit of the finite representability of
TMs. However, it is an easy exercice to implement the counter directly
into the QTM, and in the following we shall briefly describe two ways
to do it. Nevertheless, we stress that none of these implementations
can be seen as natural or standard, and that, on the other hand, one
could completely ignore the problem, assuming to add a clock to
implement the counter. For these reasons, in the paper, we have
preferred to give the more abstract solution, instead of any more
concrete implementation.

\subsection{Extra symbols}
\label{ssec:extra-symbols}

A first possibility, that we followed in a previous version of the
paper, is to duplicate the symbol alphabet $\Sigma$ by adding a set of
\emph{extra tape symbols}
$\abextra{\Sigma}=\{\symextra{a} \mid a\in\Sigma\}$: a new symbol
$\symextra{a}$, for any $a\in\Sigma$ (including the blank $\Box$). In
this way, when in a final state, the machine replaces any symbol $a$
with the corresponding extra symbol $\symextra{a}$, and moves to the
right. Dually, when in a source state, if the current cell contains an
extra symbol $\symextra{a}$, the machine replaces the current symbol
with the corresponding symbol $a$ and moves to the right; otherwise,
when the current symbol is $a\in\Sigma$, it behaves as specified by
the main transition function $\delta_0$. The legal configurations are
then restricted to three possible cases:
\begin{enumerate}
\item $\langle \alpha, q, \beta, i\rangle$
\item $\langle \alpha\symextra{\gamma}, q_t, \beta, i\rangle$
\item $\langle \alpha, q_s, \symextra{\gamma}\beta, i\rangle$
\end{enumerate}
where: $\alpha\beta\gamma\in\Sigma^*$,
$\symextra{\gamma}\in\abextra{\Sigma}^*$ is the sequence of extra
symbols obtained by replacing any symbol $a$ of $\gamma$ with the
corresponding extra symbol $\symextra{a}$; $q$ is any state; $q_t$ is
a target state; $q_s$ is a source state. It is readily seen that, to
obtain an isomorphism between QTMs with extra symbols and QTMs with
counters, it suffices to take the following bijection of
configurations (where we use the same symbols as above):
\begin{enumerate}
\item
  $\langle \alpha, q, \beta, i\rangle \mapsto \langle \alpha, q, \beta, i, 
  0\rangle$
\item
  $\langle \alpha\symextra{\gamma}, q_t, \beta,i\rangle \mapsto \langle
  \alpha, q_t, \gamma\beta, i-|\gamma|, |\gamma|\rangle$
\item 
  $\langle \alpha, q_s, \symextra{\gamma}\beta,i\rangle \mapsto \langle
  \alpha \gamma, q_s, \beta, i+|\gamma|,|\gamma|\rangle$
\end{enumerate}
where $|\gamma|$ denotes the length of $\gamma$.

\subsection{Additional counter tape}
\label{ssec:counter-tape}

\begin{figure}[!htb]
  \begin{center}
    \scalebox{0.8}{%
      \includegraphics{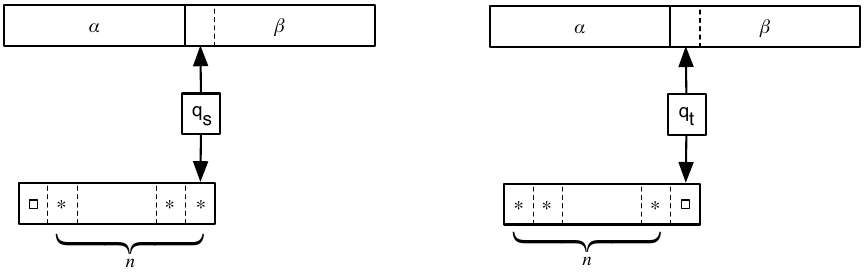} 
    }
  \end{center}
  \caption{Implementation of the counter by means of a second tape
    with only one non blank symbol $*$.}
  \label{fig:counter}
\end{figure}

Another possibility is to add a second tape to the machine. The
alphabet of this counter tape contains only one symbol $*$, in addition
to the blank $\Box$ corresponding to the empty cell. A counter
containing a value of $n$ corresponds then to a tape with $n$ symbols $*$,
see Figure~\ref{fig:counter}.

Adding/subtracting $1$ to the counter corresponds to write/delete a
$*$ symbol. To implement these operations by a single step, it
suffices that:
\begin{enumerate}
\item When the machine is in a source state $q_s$, the counter head is
  on the rightmost $*$ of the counter tape, if $n > 0$, or on an empty
  cell, if $n=0$. If the current counter symbol is a $*$, any
  transition in the state $q_s$ replaces such a $*$ with a $\Box$, and
  moves the counter head to the left, until the current counter symbol
  becomes a $\Box$, in which case the machine starts its main
  evolution.
\item When the machine is in a target state $q_t$, the counter head is
  on the first empty cell of the counter tape to the right of the
  sequence of $*$. Any transition in the state $q_t$ replaces then
  the $\Box$ in the current counter cell with a $*$, and moves the
  counter head to the right.
\item When the state is neither a source nor a target state, the
  counter tape is empty, and any transition leaves the counter tape
  unchanged.
\end{enumerate}

\end{document}